\documentclass[5p,final]{elsarticle}
\usepackage{amsfonts}
\usepackage{amssymb}
\usepackage{amsmath}  
\usepackage{amsthm}
\usepackage{enumitem} 
\usepackage{geometry}
\usepackage{graphicx}
\usepackage{algorithm}
\usepackage{algcompatible}
\usepackage{algpseudocode}
\usepackage[colorlinks=true, allcolors=blue]{hyperref}
\usepackage{hyperref}
\usepackage[utf8]{inputenc} 
\usepackage[english]{babel} 
\usepackage{url}
\usepackage{breakurl}
\usepackage{wrapfig}
\usepackage{graphicx}
\usepackage{subfiles}
\usepackage{xfrac}
\usepackage{multirow}
\usepackage{cleveref}
\usepackage{balance}
\usepackage{svg}
\usepackage{booktabs}
\usepackage{tikz}
\usetikzlibrary{shapes.geometric, arrows}
\usepackage{tikz-3dplot}
\usepackage[squaren]{SIunits}
\usepackage{caption}
\usepackage{widetext}
\usepackage[normalem]{ulem}
\usepackage{subcaption}
\usepackage{lipsum}
\usepackage{todonotes}
\usepackage{float}

\newcommand{\bc}{\begin{center}}
\newcommand{\ec}{\end{center}}
\newcommand{\be}{\begin{equation}}
\newcommand{\ee}{\end{equation}}
\newcommand{\bea}{\begin{eqnarray}}
\newcommand{\eea}{\end{eqnarray}}
\newcommand{\beq}{\begin{eqnarray*}}
\newcommand{\eeq}{\end{eqnarray*}}
\newcommand{\bv}{\left( \begin{array}{c} }
\newcommand{\ev}{\end{array} \right) }

\newcommand{\mlm}{\lim_{\substack{\Delta x \rightarrow 0 \\ \Delta t \rightarrow 0}}}

\usepackage{natbib}
\journal{arXiv.org}
\begin{document}
\title{Correlation emergence in two coupled simulated limit order books}
\author[uct-sta]{Dominic Bauer}
\ead{dominic.bauer@alumni.uct.ac.za}
\author[uct-sta]{Derick Diana}
\ead{derick.diana@alumni.uct.ac.za}
\author[uct-sta]{Tim Gebbie}
\ead{tim.gebbie@uct.ac.za}
\address[uct-sta]{Department of Statistical Sciences, University of Cape Town, Rondebosch, South Africa}

\begin{abstract}
We use random walks to simulate the fluid limit of two coupled diffusive limit order books to model correlation emergence. The model implements the arrival, cancellation and diffusion of orders coupled by a pairs trader profiting from the mean-reversion between the two order books in the fluid limit for a Lit order book with vanishing boundary conditions and order volume conservation. We are able to demonstrate the recovery of an Epps effect from this. We discuss how various stylised facts depend on the model parameters and the numerical scheme and discuss the various strengths and weaknesses of the approach. We demonstrate how the Epps effect depends on different choices of time and price discretisation. This shows how an Epps effect can emerge without recourse to market microstructure noise relative to a latent model but can rather be viewed as an emergent property arising from trader interactions in a world of asynchronous events.
\end{abstract}

\begin{keyword}
limit order book \sep  Epps effect \sep reaction-diffusion \sep non-uniform sampling
\PACS 89.65.Gh \sep 02.50.Ey
\MSC 91-04 \sep 91G10 \sep 91G80
\JEL G11 G14 G17 055
\end{keyword}

\maketitle
\newpage

\section{Introduction} \label{sec:introduction}

Financial markets operate through intricately coupled collections of order books. Order books can be differentiated at the most coarse level into either lit or latent order books, both play pivotal roles in the dynamics of price emergence. Lit order books provide an environment where all market participants can readily view and engage with displayed orders \citep{ohara2015} and actively transact. This transparency not only fosters liquidity but also enables traders to promptly identify available orders, thereby contributing to the creation of a robust and efficient market.

A distinctive attribute of lit order books is their immediate and transparent execution mechanism. Trades within lit order books transpire swiftly at the best available price, aligning with the price-time priority principle \citep{ohara2015}. In contrast, latent order books  \citep{toth2011} encompass hidden or non-displayed orders, providing a realm of confidentiality for traders, but really represent potential orders and demand provided by the largest long-term investors in a market. 

Here the obscured nature of latent order books can contribute to a reduced market impact for substantial trades, allowing market participants to execute sizeable transactions with minimised visibility and, consequently, mitigated potential influence on market prices \citep{menkveld2013}. This is often put into practice by taking large parent orders and breaking these up into many child orders that are then optimally injected into the lit order book to be executed in a manner that reduces price impact and decreases any shortfall relative to some trading target.

\citet{donier2015fully} introduced the idea of using a reaction-diffusion model of the latent order book. \citet{Gant2022b} implemented a numerical solution using a stochastic finite difference method \citep{AHJM2016, Angstmann2016} to simulate and then calibrate a reaction-diffusion market model. The model was then extended by \citet{Diana2023} to consider anomalous diffusion with non-uniform sampling times, and the code-base was developed to naturally include multiple coupled order books. We use this extended model and code base to simulate two coupled lit order books (Figure \ref{fig:price-dynamics-non-uniform}) and verify if the model can naturally generate an Epps effect as an emergent property \citep{Epps}. The Epps effect has been observed in various empirical studies of financial markets \citep{mastromatteo2010}.

We will show that this model (Equation \ref{eq:PDECoupledUpdateEquation}) produces an Epps effect when using the calibrated model parameters (Table \ref{tab:calibrated-parameters}) with non-uniform sampling (Figure \ref{fig:epps-calibrated}). The effect can be tuned to remove some of the numerical artefacts and is found without the need for market microstructure noise models relative to a preferred latent model. The Epps effect will be entirely due to discretisation and sampling supporting the arguments put forward by \citet{chang2021using} where it is warned that so-called market-microstructure noise models may merely be the {\it ad hoc} models required to defend the use of particular latent models, rather than capturing foundational empirical features of real financial markets. That the model representation can easily be confused with the underlying market reality itself -- leading to possibly erroneous risk management and regulatory decisions because global in-the-limit properties can then be conflated with local data and sample-specific properties under the averaging required to link theory to data.

\section{The coupled limit order books} \label{sec:2}

\subsection{Coupled partial differential equations} \label{ssec:couplings}

We assume that the order imbalance drives trading and start with an order survival function $\theta$ that describes the accumulated volume of orders that have not been removed by trading up to some time $t$ since the start of trading. The survival function can be written in terms of some order removal rate $a(x,t)$ which is a function of the log-price $x$ and some calendar time $t$ \cite{Diana2023}:
\begin{equation}
    \theta = \int_0^t a(x,\tau) d\tau. 
\end{equation}
Following \citet{Diana2023} we can consider a source term $c(x,t)$ that captures the creation of orders, and then assume that orders diffuse in the order book as a function of price through time via anomalous diffusion. The anomalous diffusion can be parameterised using two parameters: first, the anomalous diffusion parameter $D_{\alpha}$, and second, a fractional time scaling parameter $\alpha$. The diffusion can then be generated by a Riemann-Liouville operator $D_t^{1-\alpha}$ for times $t$.

We can then define two (or more) order book equations $j \in \{ 1,2 \}$ in the presence of information shocks that can be synchronised (or unsynchronised) where the reaction-diffusion equation describing the time evolution of the $j^{th}$ order books density $\varphi^{(j)}(x,t)$ is:
\begin{align}
\varphi^{(j)}_t =&
D^{(j)}_\alpha \left[\theta^{(j)} D^{1-\alpha}_{t} \left( \frac{\varphi^{(j)}}{\theta^{(j)}} \right)\right]_{xx} \negthickspace  \negthickspace  \negthickspace \nonumber \\ &+ V_t^{(j)} \left[\theta^{(j)} D^{1-\alpha}_{t} \left( \frac{\varphi^{(j)}}{\theta^{(j)}} \right)\right]_x \negthickspace  \negthickspace - a^{(j)} \varphi^{(j)} + c^{(j)}.  \label{eq:coupledRDequation}
\end{align}
Here $j$ is the order book index so that, for example, $D^{(j)}_{\alpha}$ is the order diffusion rates for the $j^{th}$ limit order book. Similarly, for the source terms $c^{(j)}$, the removal rates $a^{(j)}$, and the associated survival functions $\theta^{(j)}$. 

The random driving forces are $V_t^{(j)}$. The driving forces $V_t^{(j)}$ are taken to be a Brownian Motions. Here the information shocks are synchronised and there is only one source for the stochastic force potential $V_t$, and we drop the $j$ index.

The trade prices for the $j^{th}$ order book are:
\begin{equation}
p^{(j)}(t) = {{x} \,\, \operatorname{s.t.}\,\,  \varphi^{(j)}(x,t)=0}.
\end{equation}
We assume that the annihilation rates are constants:
\begin{equation}
a^{(j)}(x,t)=\nu_j.
\end{equation}
The creation term is separated further into a source terms $s^{(j)}$, coupling terms $\ell^{(j,k)}$, and shocks $\delta^{(j)}$ for the $j^{th}$ assets order book:
\begin{equation}
c^{(j,k)}(x,t)=s^{(j)}(x,t)+\ell^{(j,k)}(x,t)+ \delta^{(j)}(x,t).
\end{equation}
Order book shocks $\delta^{(j)}$ are used to estimate the price impact \citep{Diana2023}. These involve introducing a shock of size $Q$ and measuring the change in price as a result. The source terms $s^{(j)}$ are assumed to be either latent order book sources or lit order book sources. Lit order book source terms have vanishing boundary conditions, while the latent order book have finite boundary conditions values. 

If there is an pair-wise order book coupling from the $j^{th}$ order book to (say) the $k^{th}$ order book this will be carried by the coupling term $\ell^{(j,k)}$. If there is no coupling then we can drop the $k$ index on the source term. Boundary condition considerations (to ensure that that the integral under $\varphi^{(i)}$ is constant) make it convenient that we will only consider the lit order book sources \citep{Gant2022b,Diana2023}:
\begin{equation}\label{eq:source-term}
s^{(j)}(x,t) = - \lambda_j \mu_j (x - p^{(j)}(t)) e^{-\mu_j\left(x - p^{(j)}(t)\right)^2}. \\
\end{equation}
The Julia implementation uses the equivalent decaying-Gaussian parameterisation $e^{-(\mu y)^2}$ \citep{bauer2024-zivahub,bauer2024-github}.
Here, using pair-wise couplings between the $j^{th}$ and $k^{th}$ order books \cite{diana2023-zivahub}:
\begin{equation}
\begin{aligned} 
    \ell^{(j,k)}(x,t)=G(x,t,p^{(j)},\Delta p_{jk}) = G_j(x,t) \\
    \ell^{(k,j)}(x,t)=G(x,t,p^{(k)},\Delta p_{kj}) = G_k(x,t). \label{eq:ell-crosscoupling} 
\end{aligned}
\end{equation}
The coupling equations are a function of the difference in mid-prices of two order books where, if $p^{(j)}$ (in $\varphi^{(j)}$) is above $p^{(k)}$ (in $\varphi^{(k)}$), then more bids are placed above the mid-price in $\varphi^{(j)}$ to push the mid-price down to $\varphi^{(k)}$'s mid-price. To achieve this we define:
\begin{equation}
g_j(x,t) = - \lambda_j \mu_j x e^{-\mu_j x^2}.
\end{equation} 
Here $\lambda_j$ and $\mu_j$ are constants specific to the $j^{th}$ order book, and $\Delta p_{jk}$ is the difference between the mid-prices of the two order books: $\Delta p_{jk} = p^{(j)} - p^{(k)}$:
\begin{equation}
\begin{aligned}
    G_j = 
   \left\{
    \begin{array}{ll}
        g_j\left(x-p^{(j)}(t)\right)\, \Delta p_{jk} &,x > p^{(j)}(t), \Delta p_{jk} > 0\\
        g_j\left(\frac{j}{\Delta p_{jk}}(x-p^{(j)}(t))\right) &,x \leq p^{(j)}(t), \Delta p_{jk} > 0 \\
        g_j\left(x-p^{(j)}(t)\right)\, \Delta p_{jk} &,x \leq p^{(j)}(t), \Delta p_{jk} \leq 0\\
        g_j\left(\frac{1}{\Delta p_{jk}}(x-p^{(j)}(t))\right) &,x > p^{(j)}(t), \Delta p_{jk} \leq 0.
    \end{array}
    \right.  \label{eq:G-coupling}
\end{aligned} \nonumber
\end{equation}
The Julia implementation uses bounded side-dependent rescaling; see \citep{bauer2024-zivahub,bauer2024-github}.
This has the interpretation of an external agent (such as a pairs trader) observing the system, and buying (or selling) one asset according to whether the mid-price of the other asset is above (or below) some price threshold. In this way, the system can be generalised to many assets being traded by pair traders. Here we will focus on only two assets. 

\subsection{Lattice parameters} \label{sssec:DtDx}

We assume that the diffusion limit exists 
\begin{equation}\label{eq:d}
    D_{\alpha} = \mlm  \frac{r}{2}\frac{\Delta x^2}{\Delta t^{\alpha}},
\end{equation}
where $D_{\alpha}$ is some diffusion parameter. We can then use this to set the lattice price grid increments in terms of the lattice time increments following \citet{Diana2023}. We will treat this parameter as the same across all the order books, similarly for the fractional time parameter $\alpha$.

To simulate non-uniformly sampled grids in time we then assume that the time changes are exponentially distributed: $\Delta t \sim \mathrm{Exp}(\lambda_{i} t)$. Using this, we can have two different clocks, one for each of the two order books as defined by the two intensities, $\lambda_1$ and $\lambda_2$ \citep{Diana2023} (Table \ref{tab:parameters}).

\subsection{Limit order book parameters} \label{ssec:lobparams}

Table \ref{tab:parameters} gives the model parameters. $L$ is the system length and $M$ is the number of divisions within the system length. Both are used to determine the log-price grid size of $\Delta x$ where $\Delta x = \frac{L}{M}$. $D_{\alpha}$ is the diffusion rate described in Equation [\ref{eq:d}], $r$ is the probability of a jump occurring (which would lead a change in price), $\nu$ is the rate at which orders in the order book are cancelled, $\alpha$ is the fraction of the derivative (in Equation [\ref{eq:d}]). The initial prices for the two order books are $p_{1,2}(0)$, where $\lambda_{1,2}$ and $\mu_{1,2}$ are the source term variables defined in Equation [\ref{eq:source-term}] for the two order books. 

\subsection{Coupled update equations} \label{ssec:coupledPDE}

The coupled simple diffusion equation given in Equation [\ref{eq:coupledRDequation}] can now be numerically solved using non-uniformly sampled update equations \citep{Diana2023} for the set of coupled order book equations. With some approximation function $\varphi^{(j)}_{\Delta}(x,t)$ for the order book densities on some background lattice \cite{Diana2023} so that $\varphi^{i(j)}_n = \varphi^{(j)}_{\Delta}(x_i,t_n)$ where the grid spacing (in the background lattice) are uniform $\Delta x$ but the times are non-uniformly sampled increments $\Delta t_n$. At some time $t_{n-1}$ and with some time increment $\Delta t_{n-1} = t_{n} - t_{n-1}$ we can then use the diffusion constraint in Equation [\ref{eq:d}] to find the unique price increment at time $t_{n-1}$, {\it i.e.} $\Delta x_{n-1}$. Then we can find the prices from which the right and left jumps will occur that are consistent with the diffusion to the order of the approximations:
\begin{equation}
 \hat \varphi^{i\pm 1(j)}_{m} = \varphi^{(j)}_{\Delta}(x^{i \pm 1}_{m}, t_{m}) = \varphi^{(j)}_{\Delta}(x_i \pm \Delta x_{m}, t_{m}).
\end{equation}
The probabilities of left and right jumps do not depend on the sequence $\left\{\Delta x_m\right\}_{m=1}^M$, they only depend on the most recent entry at $n-1$. In contrast, $\varphi^{(j)}$ makes use of the entire history of the sequence $\left\{\Delta x_m\right\}_{m=1}^M$ where at each time $t_m$ we have the unique $\Delta t_m$ and hence its unique $\Delta x_m$ relative to the background points $x_i$. We can then find the appropriate non-uniformly sampled update equation describing the evolution of order for each order book \citep{Diana2023}:
\begin{widetext}
\begin{equation}
\varphi^{i(j)}_n = \sum_{m=0}^{n-1} K_{n-m} e^{-\nu (t_{n-1}-t_m)} \left[{\tfrac{1}{2}\left({r+F^{(j)}_{n-1}}\right) \hat \varphi^{i-1(j)}_{m} + \tfrac{1}{2}\left({r-F^{(j)}_{n-1}}\right) \hat \varphi^{i+1(j)}_{m} - r \varphi^{i(j)}_{m} }\right] + e^{-\nu \Delta t_{n-1}} \varphi^{i(j)}_{n-1} + c^{(j,k)}_{i,n-1} \Delta t_{n-1}. \nonumber
\label{eq:PDECoupledUpdateEquation}
\end{equation}
\end{widetext}
We approximate $\hat \varphi^{i\pm 1}_{m}$ from each $x^{i \pm 1}_{n-1}$, and $x_i$ at each time $t_m$ \cite{Diana2023}. Here $K_{n-m}$ is a memory kernel for a process with Sibuya waiting times \cite{ANGSTMANN2016508} which defines the memory due to the fractional diffusion.

\section{Model Simulation}

\subsection{Limit order book parameters} \label{ssec:lobsimparams}
\begin{figure}[th]
\centering
\includegraphics[width=1\linewidth]{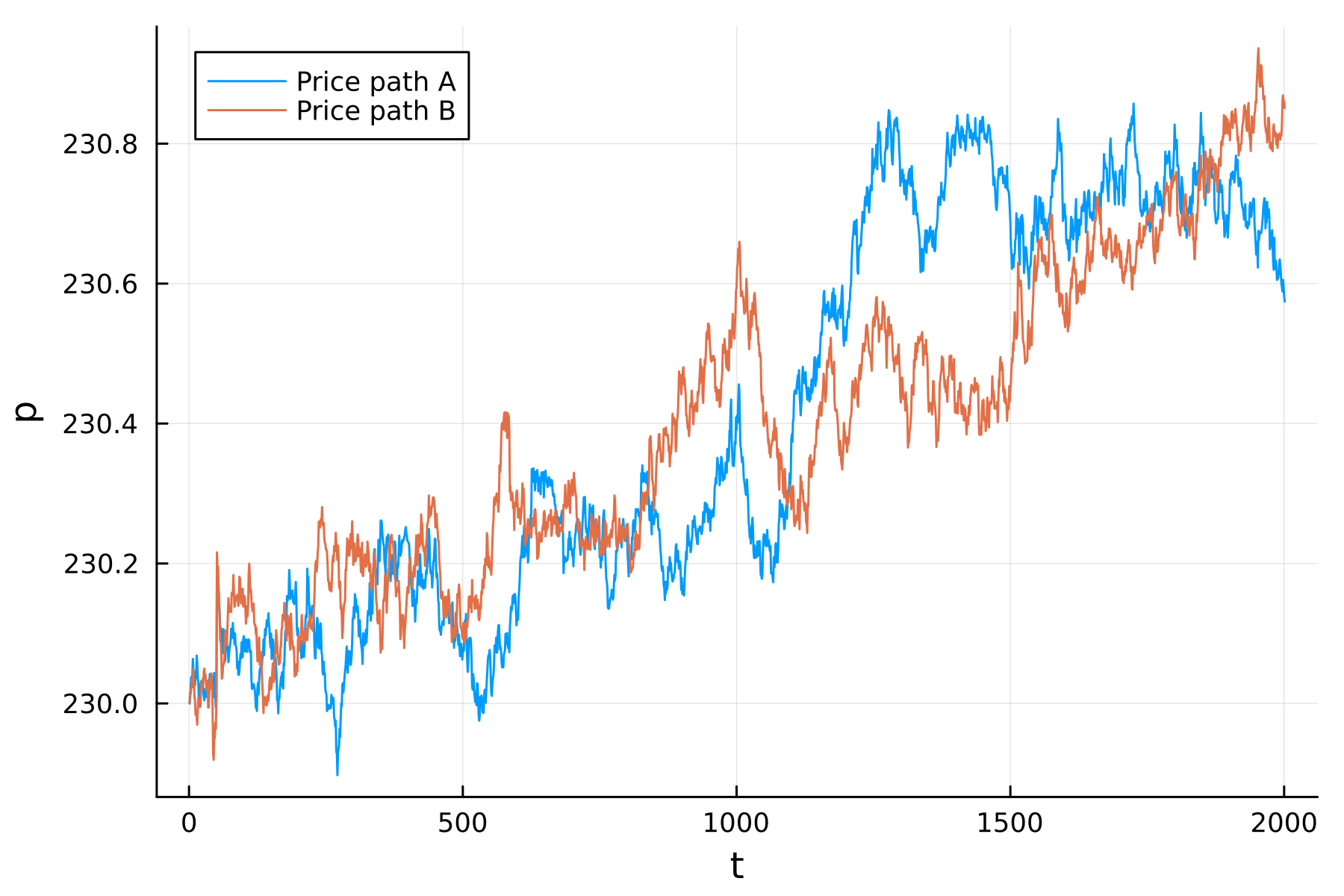}
\caption{The figure shows simulated price paths generated using the non-uniformly sampled coupled order book model given in Equation [\ref{eq:coupledRDequation}]. Here using calibrated parameter values found in Table \ref{tab:calibrated-parameters}.}
\label{fig:nonuniform-pricepaths}
\end{figure}

The parameters in Table \ref{tab:parameters} are used in the implementation by \cite{Diana2023, diana2023-zivahub} and are used in the proceeding graphs for this section. In Figure \ref{fig:nonuniform-pricepaths} we demonstrate price paths generated using our model when considering a two-coupled order book using the parameters from Table \ref{tab:parameters}. The price paths in the figures for order book A and B are blue and red, respectively. Here we will consider a single shared Diffusion parameter, cancellation rate, and fractional time parameter across all the coupled order books. 

\begin{table}[h!]
\centering
\begin{tabular}{clll} 
 \hline
Parameter & Description & Type & Value\\ 
 \hline
 $L$ & System length & Fixed & $200$ \\ 
 $M$ & Number of divisions & Fixed & $400$\\
  $r$ & Probability of self jump & Fixed & $0.5$\\ 
 $D_{\alpha}$ & Diffusion rate & Free & $0.5$\\ 
 $\nu$ & Cancellations rate & Free & $14.0$\\ 
 $\alpha$ & Fractional time & Free & $1.0$\\
 $p_{1,2}(0)$ & Initial prices & Fixed & $230.0$\\ 
 $\lambda_{1,2}$ & Source terms intensities & Fixed & $1.0$\\ 
 $\mu_{1,2}$ & Source terms rates & Fixed & $0.1$\\ 
  $\Delta x$ & 
Change in $x$ ($\frac{L}{M}$) & Fixed & $0.5$\\ 
    $\Delta t$ & Change in $t$ (Eq. [\ref{eq:d}]) & Free & $0.0625$\\ 
 \hline
\end{tabular}
\caption{The base model parameters used in the coupled order book model with accompanying description. Also shown is whether a value was fixed or free for the model calibration. $\Delta t$ is a free parameter due to it being a function of $D_{\alpha}$ which is a free parameter. These parameters are used in this section to plot figures \ref{fig:nonuniform-pricepaths},  and \ref{fig:price-dynamics-non-uniform}. They are also used in to generate the Epps effect plot (Figure \ref{fig:epps-calibrated}).} \label{tab:parameters}
\end{table} 

\subsection{Price dynamics} \label{sssec:price-paths}

\begin{figure*}[ht!]
\centering
\begin{subfigure}{0.33\textwidth}
\includegraphics[width=1\linewidth]{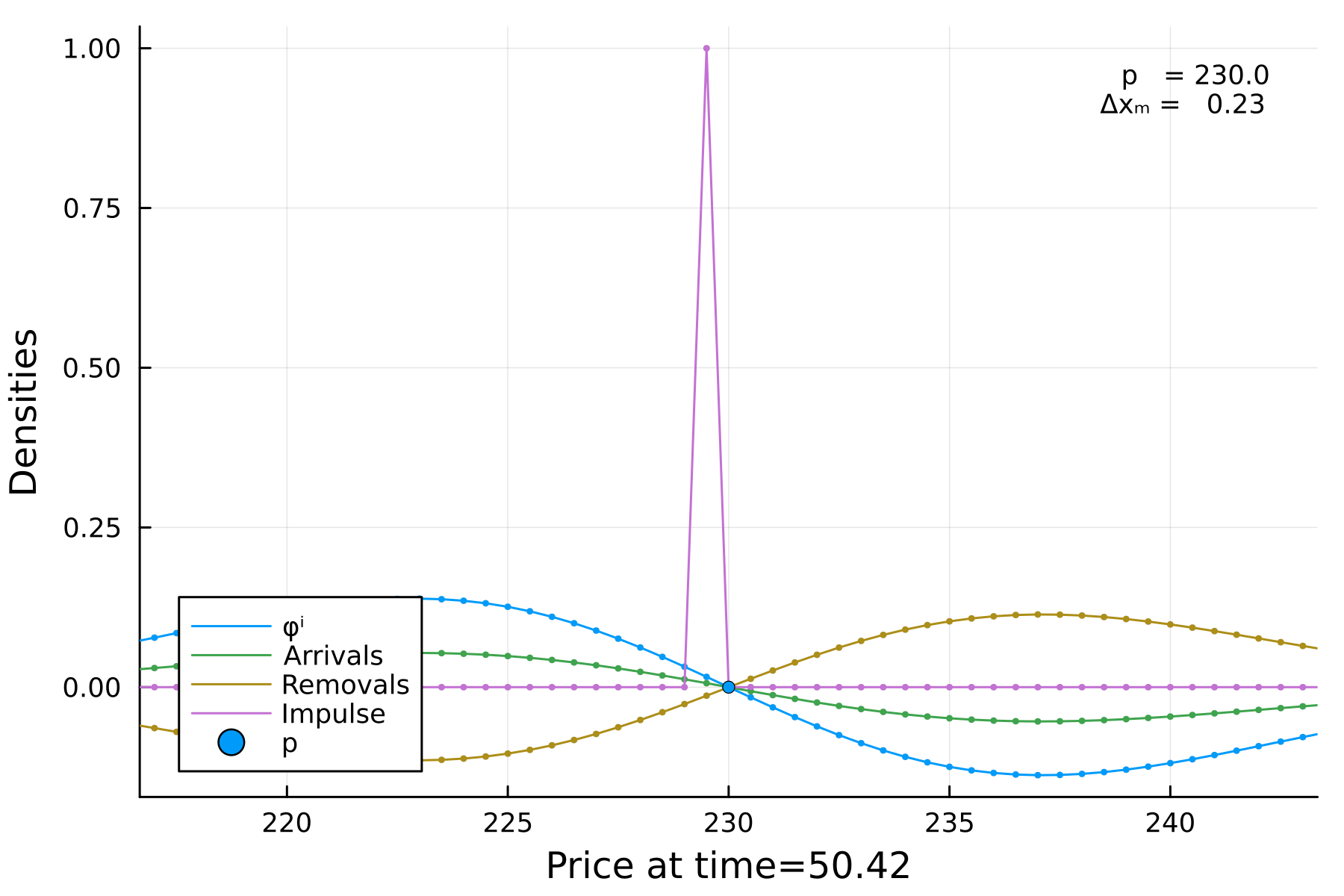}
\caption{Time 1}
\label{subfig:price-dynamics-non-uniform-1}
\end{subfigure}\hfill
\begin{subfigure}{0.33\textwidth}
\includegraphics[width=1\linewidth]{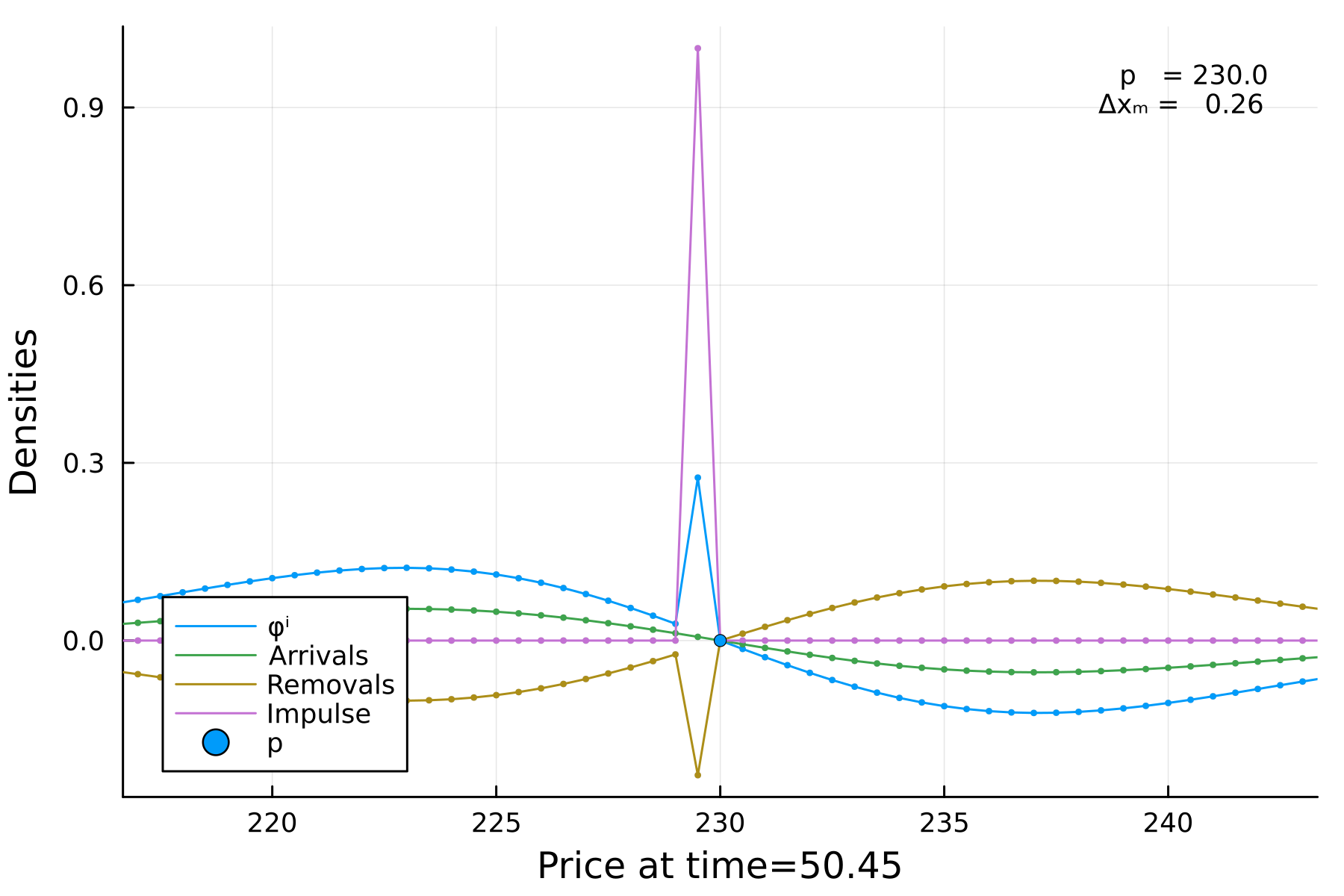}
\caption{Time 2}
\label{subfig:price-dynamics-non-uniform-2}
\end{subfigure}\hfill
\begin{subfigure}{0.33\textwidth}
\includegraphics[width=1\linewidth]{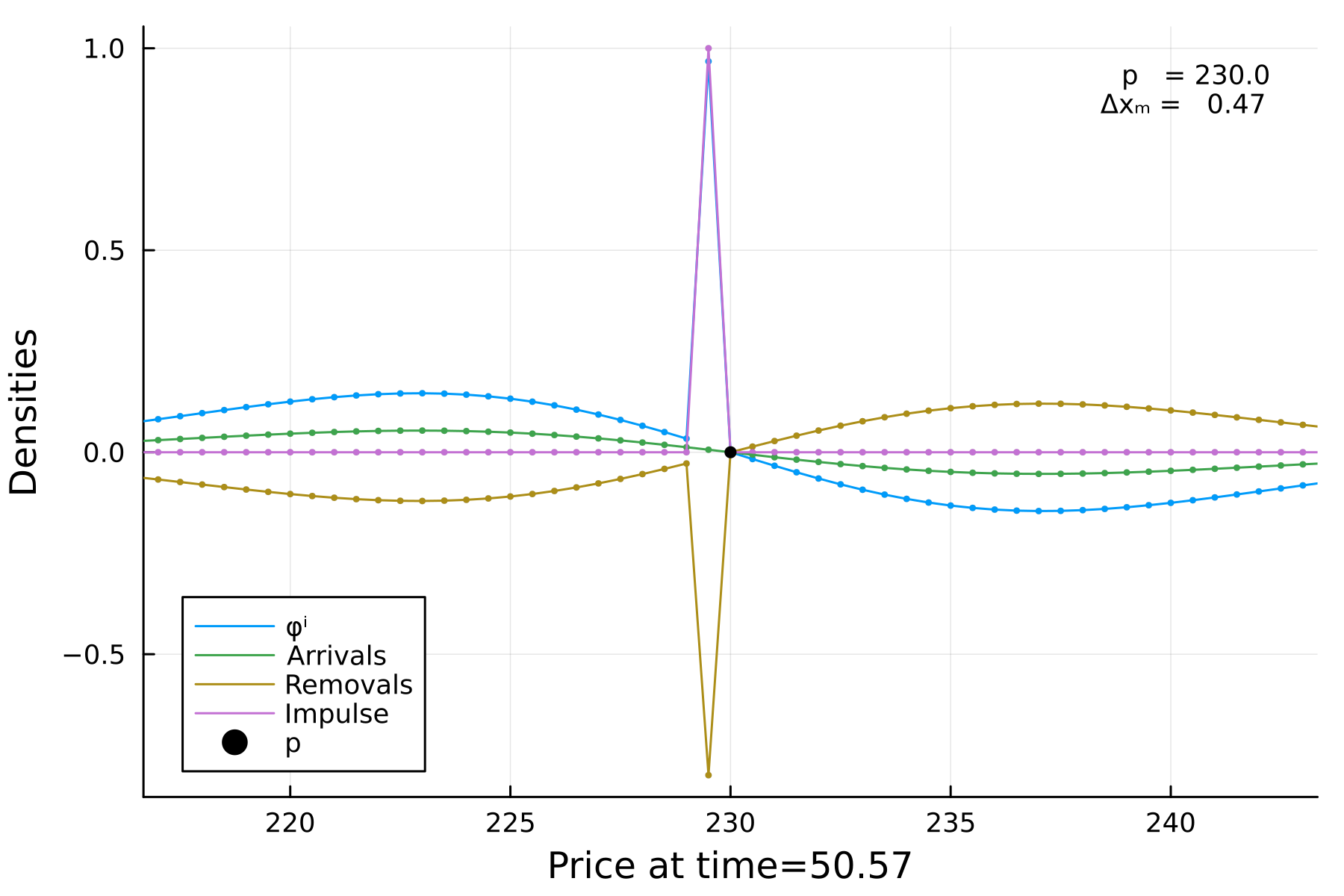}
\caption{Time 3}
\label{subfig:price-dynamics-non-uniform-3}
\end{subfigure}
\par\bigskip
\begin{subfigure}{0.33\textwidth}
\includegraphics[width=1\linewidth]{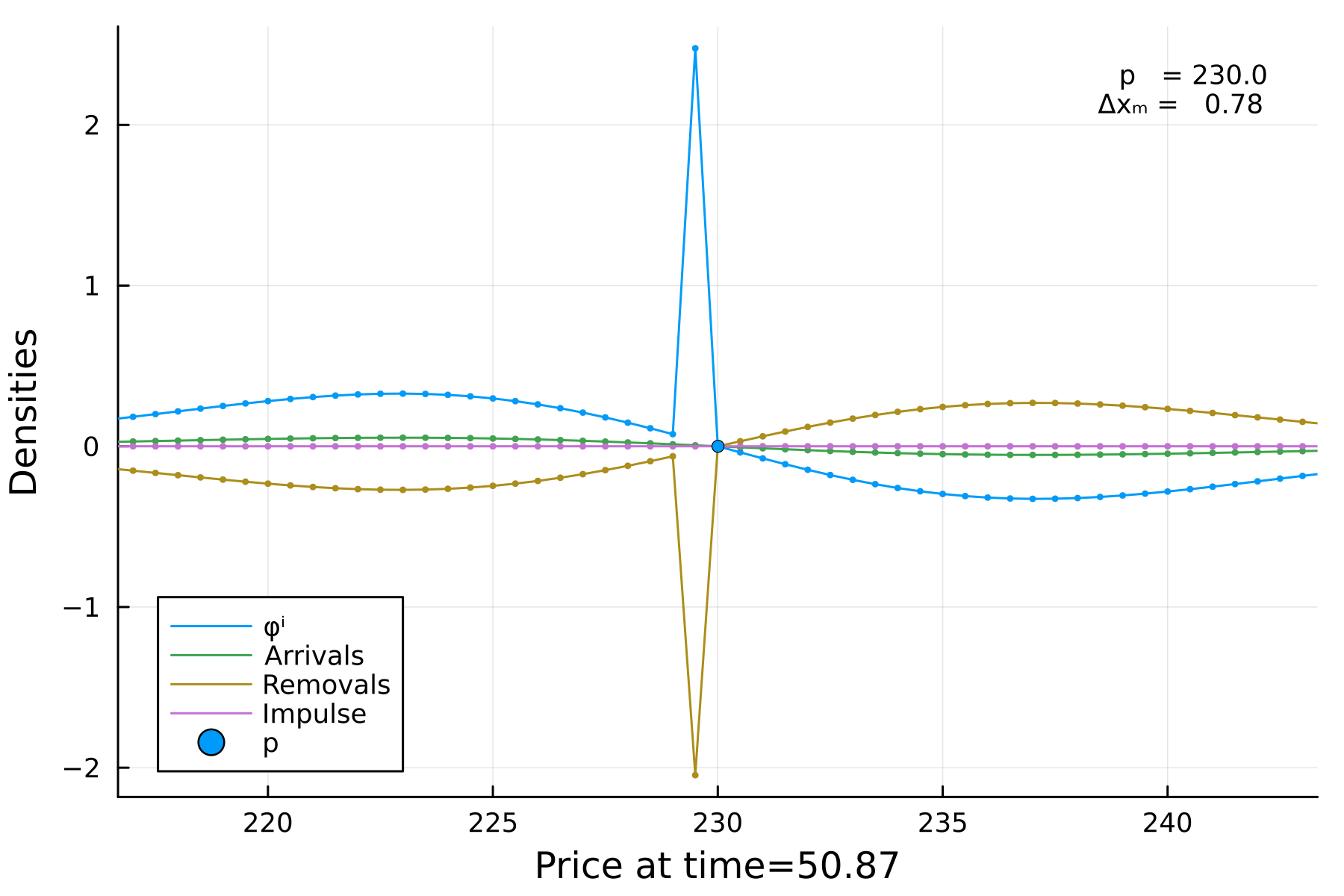}
\caption{Time 4}
\label{subfig:price-dynamics-non-uniform-4}
\end{subfigure}\hfill
\begin{subfigure}{0.33\textwidth}
\includegraphics[width=1\linewidth]{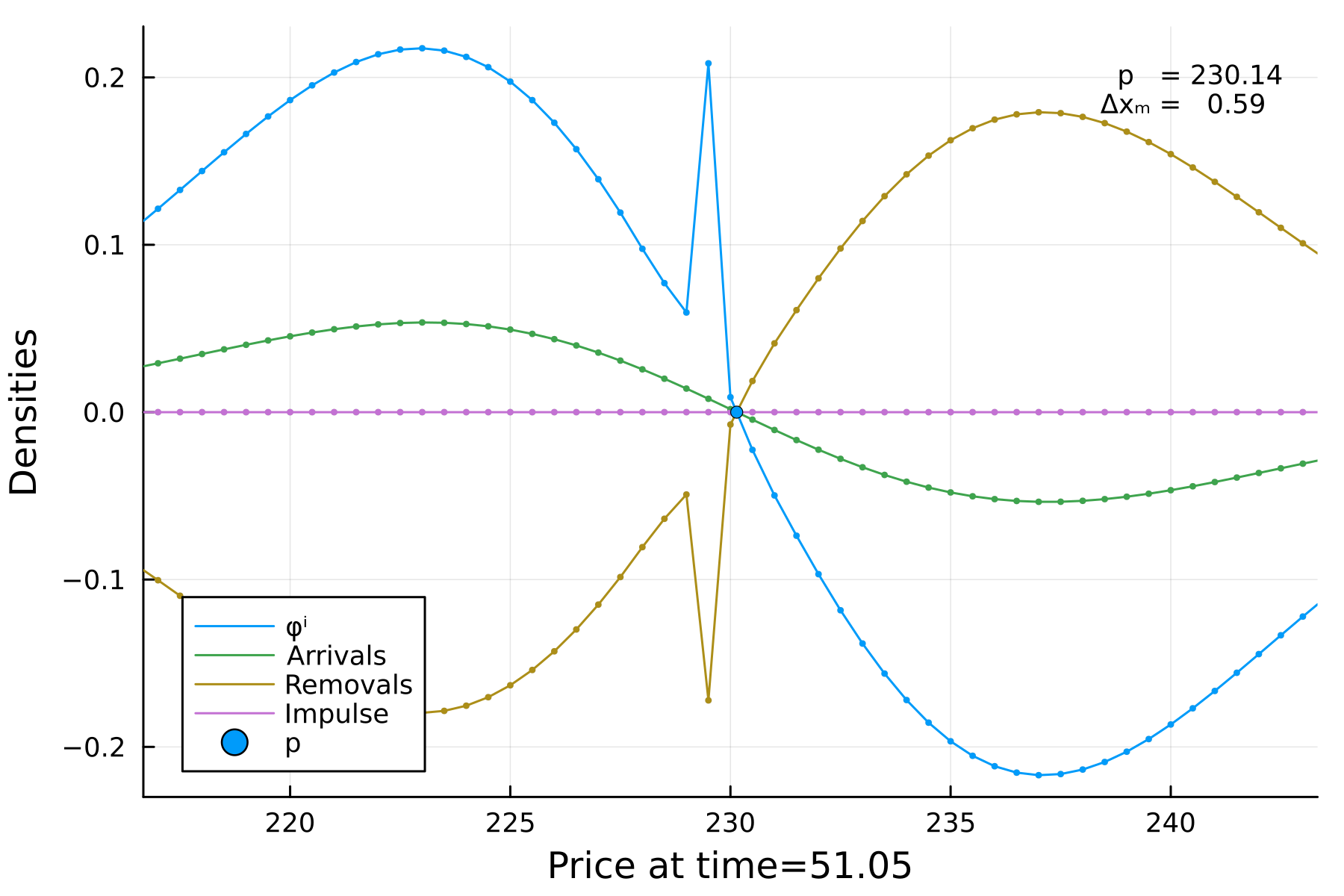}
\caption{Time 5}
\label{subfig:price-dynamics-non-uniform-5}
\end{subfigure}\hfill
\begin{subfigure}{0.33\textwidth}
\includegraphics[width=1\textwidth]{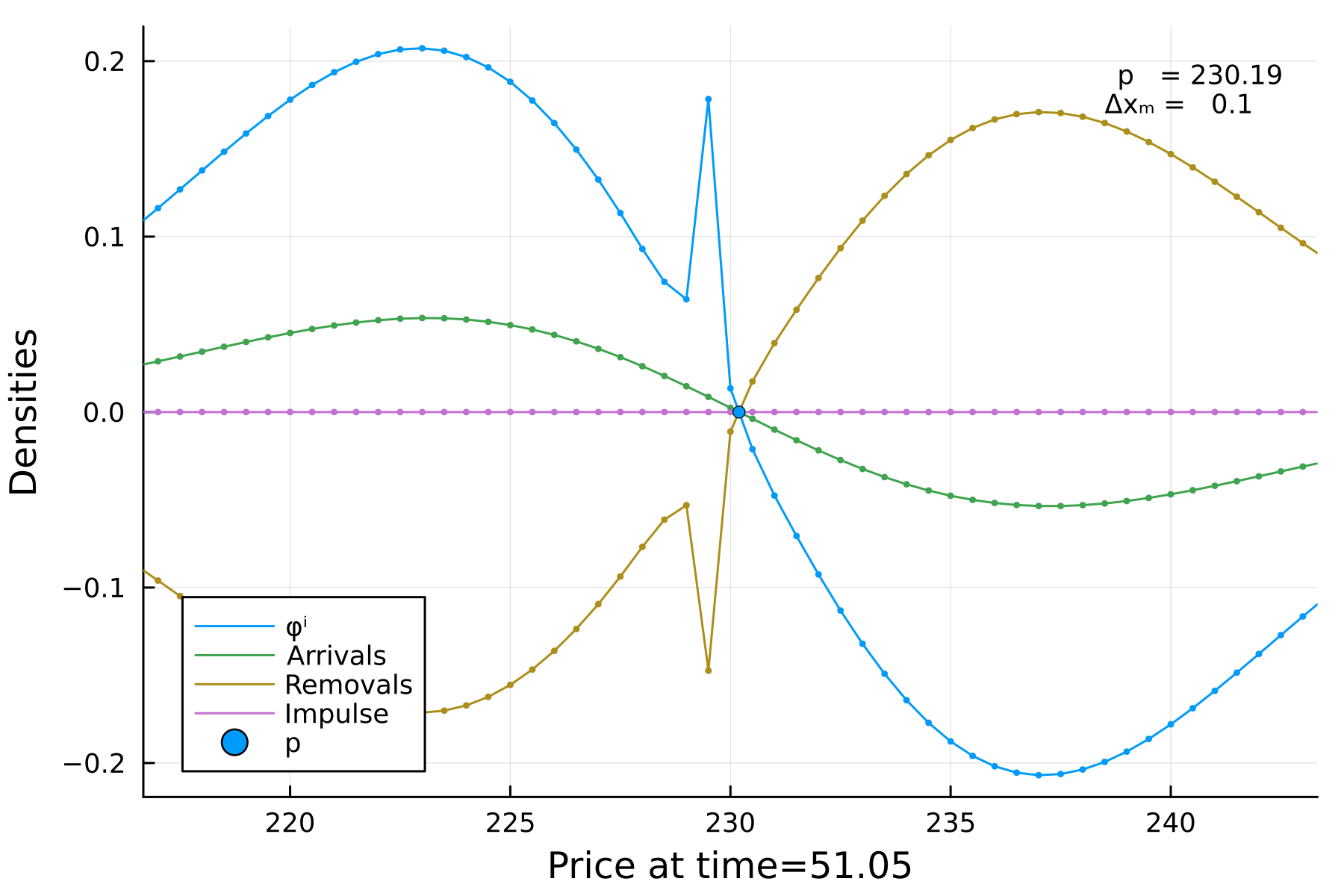}
\caption{Time 6}
\label{subfig:price-dynamics-non-uniform-6}
\end{subfigure}
\par\bigskip
\begin{subfigure}{0.33\textwidth}
\includegraphics[width=1\linewidth]{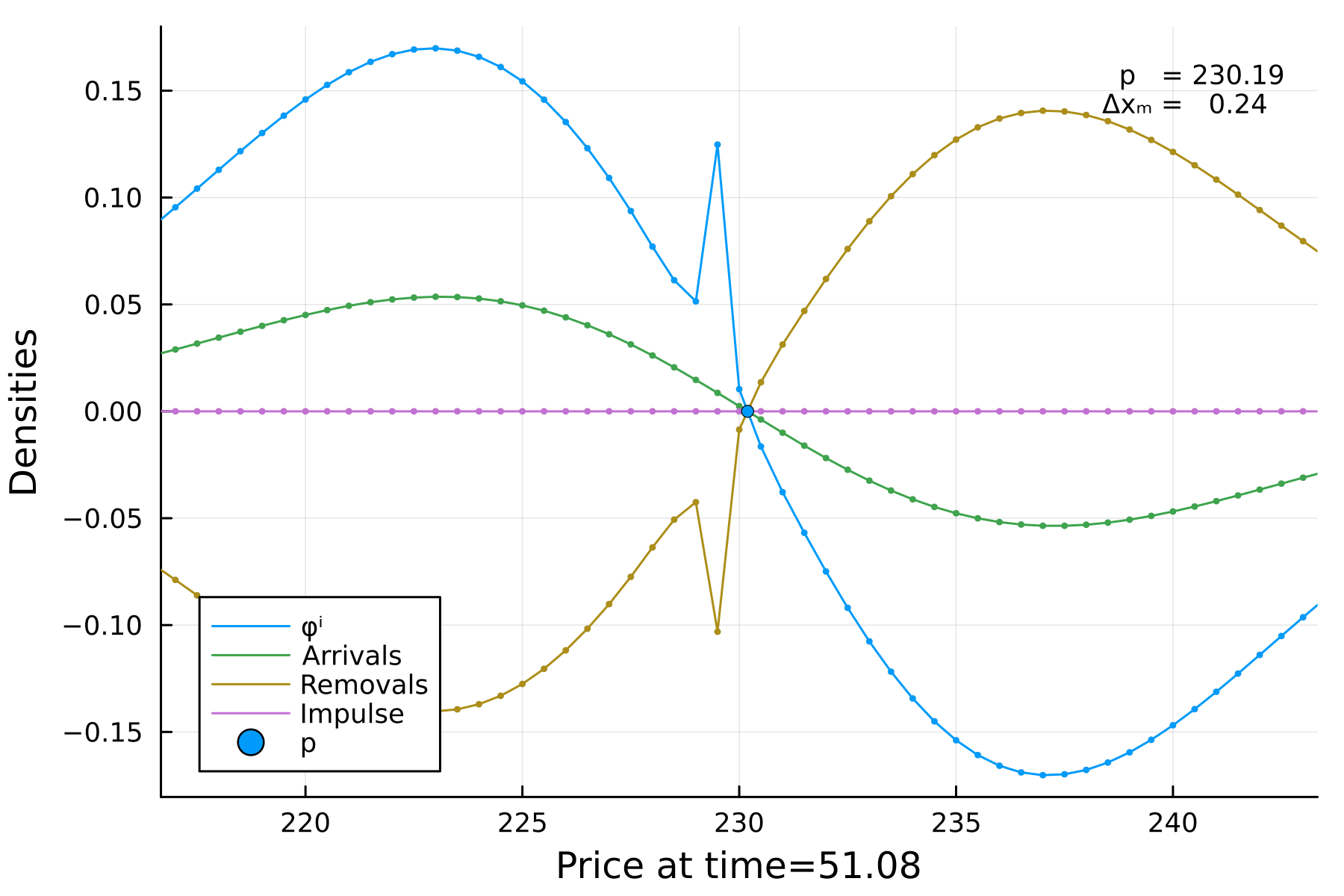}
\caption{Time 7}
\label{subfig:price-dynamics-non-uniform-7}
\end{subfigure}\hfill
\begin{subfigure}{0.33\textwidth}
\includegraphics[width=1\linewidth]{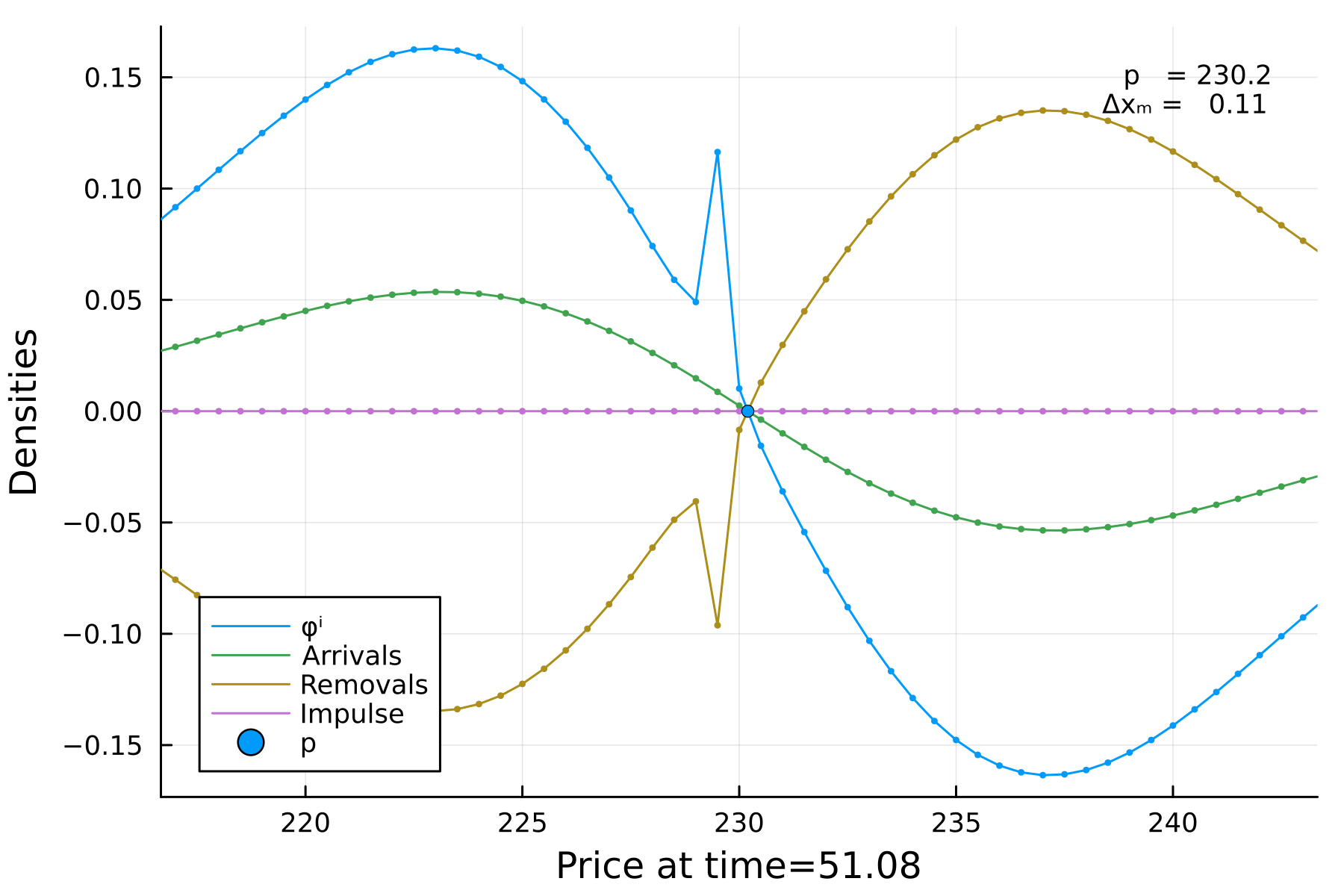}
\caption{Time 8}
\label{subfig:price-dynamics-non-uniform-8}
\end{subfigure}\hfill
\begin{subfigure}{0.33\textwidth}
\includegraphics[width=1\linewidth]{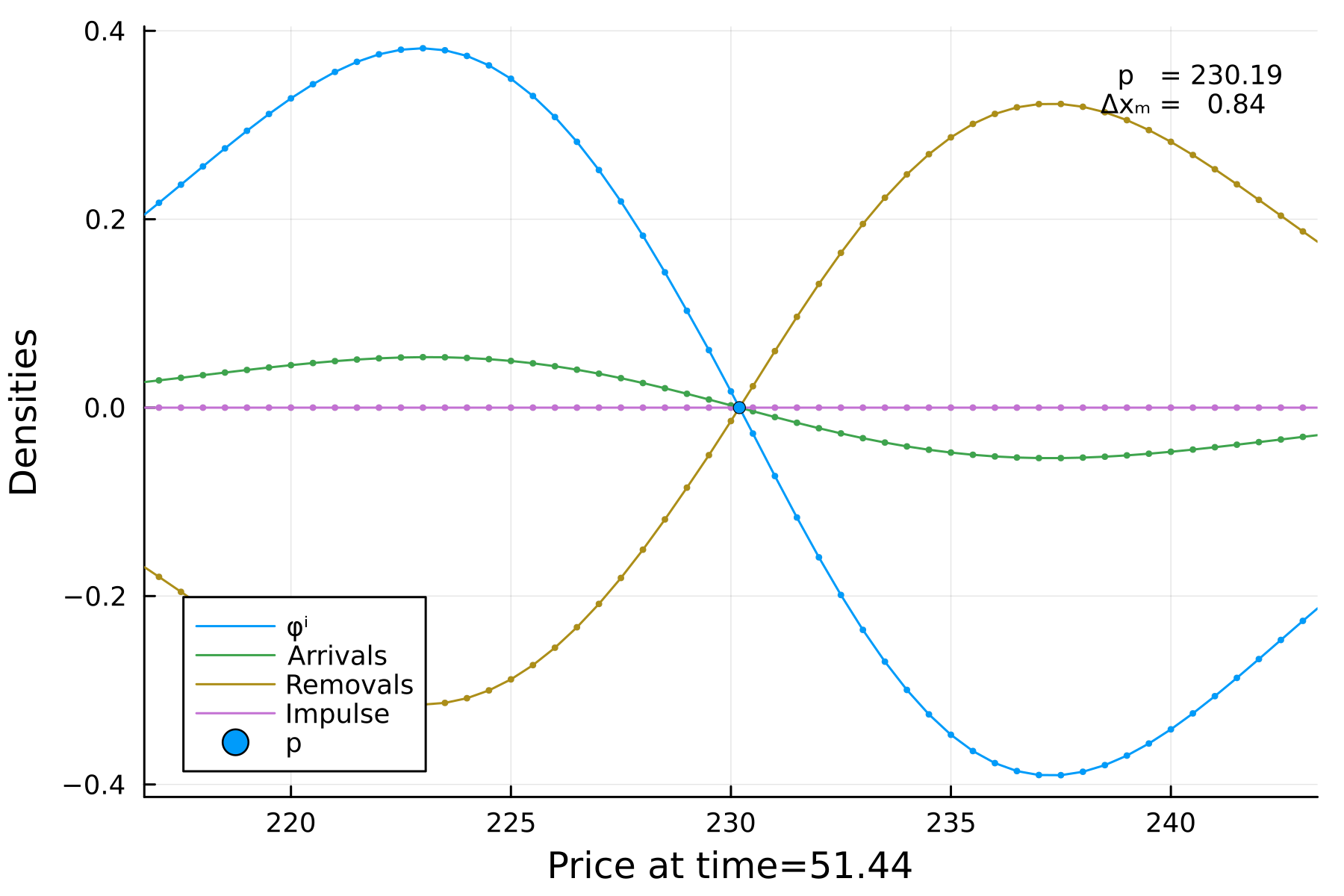}
\caption{Time 9}
\label{subfig:price-dynamics-non-uniform-9}
\end{subfigure}\hfill
\caption{Figures \ref{subfig:price-dynamics-non-uniform-1} to \ref{subfig:price-dynamics-non-uniform-9} show a sequence of snapshots depicting the effect of an order book shock on one of the order books which increases price $p$ when we use non-uniform $\Delta t$ for the coupled order book equation in Equation [\ref{ssec:coupledPDE}] with parameters defined in Table \ref{tab:parameters}. Since $\Delta t$ is varied at each time step this results in a different $\Delta x_{m}$ for each time step. $\Delta x_{m}$ and the mid-price $p$ for each snapshot are shown on the top right of each figure. We use pink to indicate an order book shock, gold to indicate order cancellations, green to indicate order arrivals and blue to indicate the mid-price. Note that some of the time increments appear the same but this is as a result of rounding. Where all the lines meet is the equilibrium and forms the mid-price $p$.}
\label{fig:price-dynamics-non-uniform}
\end{figure*}

In Figure \ref{fig:price-dynamics-non-uniform} we show snapshots of time increments $\Delta t$  of the effect an order book shock has on the mid-price $p$ when applying a non-uniform $\Delta t$ to our model, respectively. Here we are using the model parameters from Table \ref{tab:parameters}. In the subplots for both figures, blue represents the density of the order book we are modelling at the current time, green represents the orders which are about to be added via the source term, gold represents the orders which are about to be removed via the removal rate and pink represents an order book shock. The state of equilibrium in the system is where all the lines meet (i.e. the order density is $0$) and is the mid-price $p$. The parameters $p$ and $x_{m}$ are shown at the top right for each time increment $\Delta t_{m}$. 

In Figure \ref{fig:price-dynamics-non-uniform} we start with an initial order book shock as shown by the pink spike in Figure \ref{subfig:price-dynamics-non-uniform-1}. This results in an increase in density and a counteracting decrease in removals as shown in Figure \ref{subfig:price-dynamics-non-uniform-2}. This increase in density and decrease in removals persists in figures \ref{subfig:price-dynamics-non-uniform-3}, \ref{subfig:price-dynamics-non-uniform-4} and \ref{subfig:price-dynamics-non-uniform-5} at which point the shock is gone and the density and removals remain. Note that the removal density is smaller in magnitude to the density of the order book indicating that there is more willingness to buy than to sell and as such will result in a price increase to accommodate for this increased demand. We see this play out as the system seeks a new equilibrium by the increases in the mid-price, as seen in figures \ref{subfig:price-dynamics-non-uniform-6}, \ref{subfig:price-dynamics-non-uniform-7} and \ref{subfig:price-dynamics-non-uniform-8}. Finally in Figure \ref{subfig:price-dynamics-non-uniform-9} the system reaches equilibrium at an increased mid-price $p = 230.19$.

\begin{figure*}
\centering
\begin{subfigure}[t]{0.32\textwidth}
    \includegraphics[width=\textwidth]{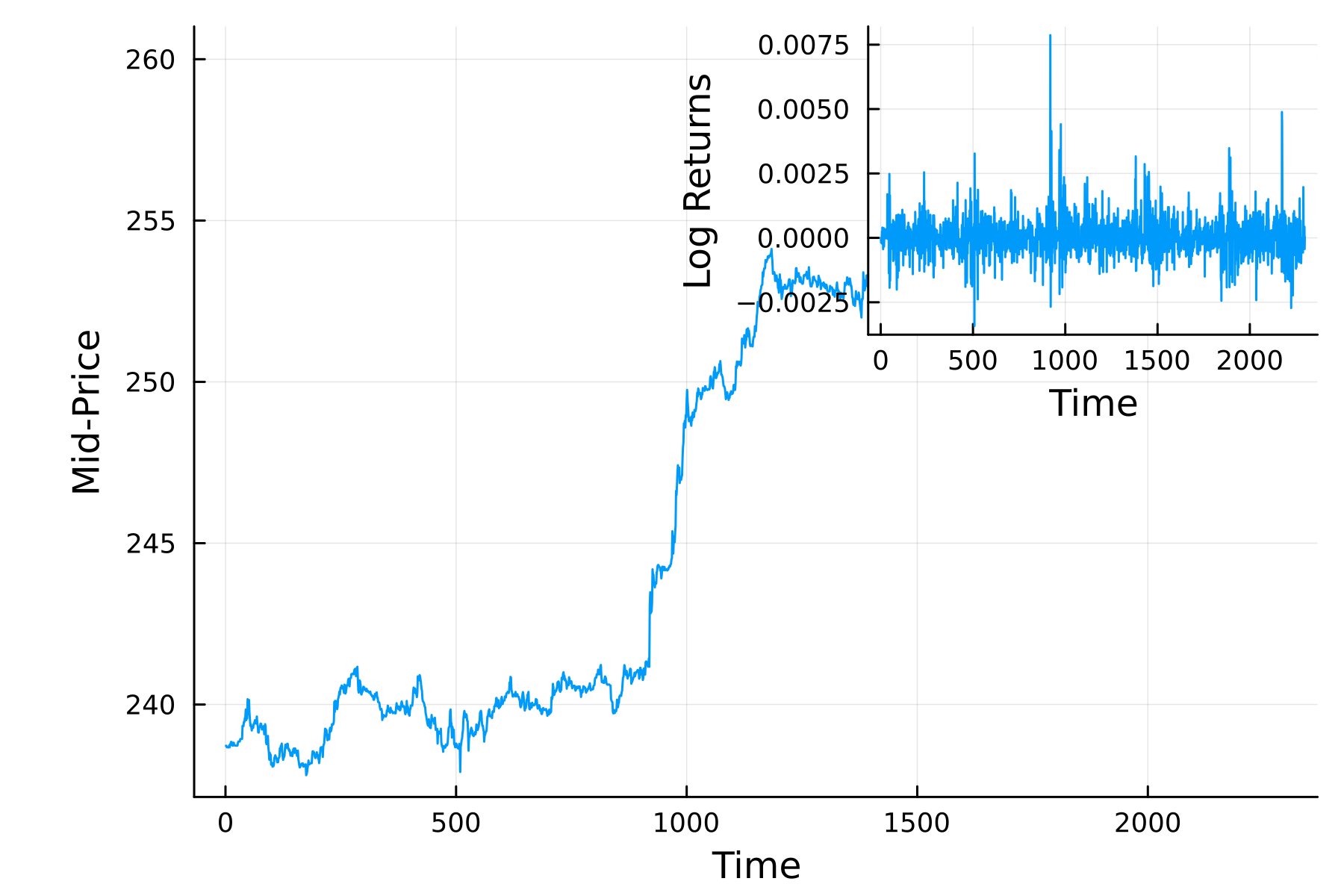}
    \caption{Empirical Price Path \& Returns} \label{subfig:sf-emp-price}
\end{subfigure}\hfill
\begin{subfigure}[t]{0.32\textwidth}
    \includegraphics[width=\textwidth]{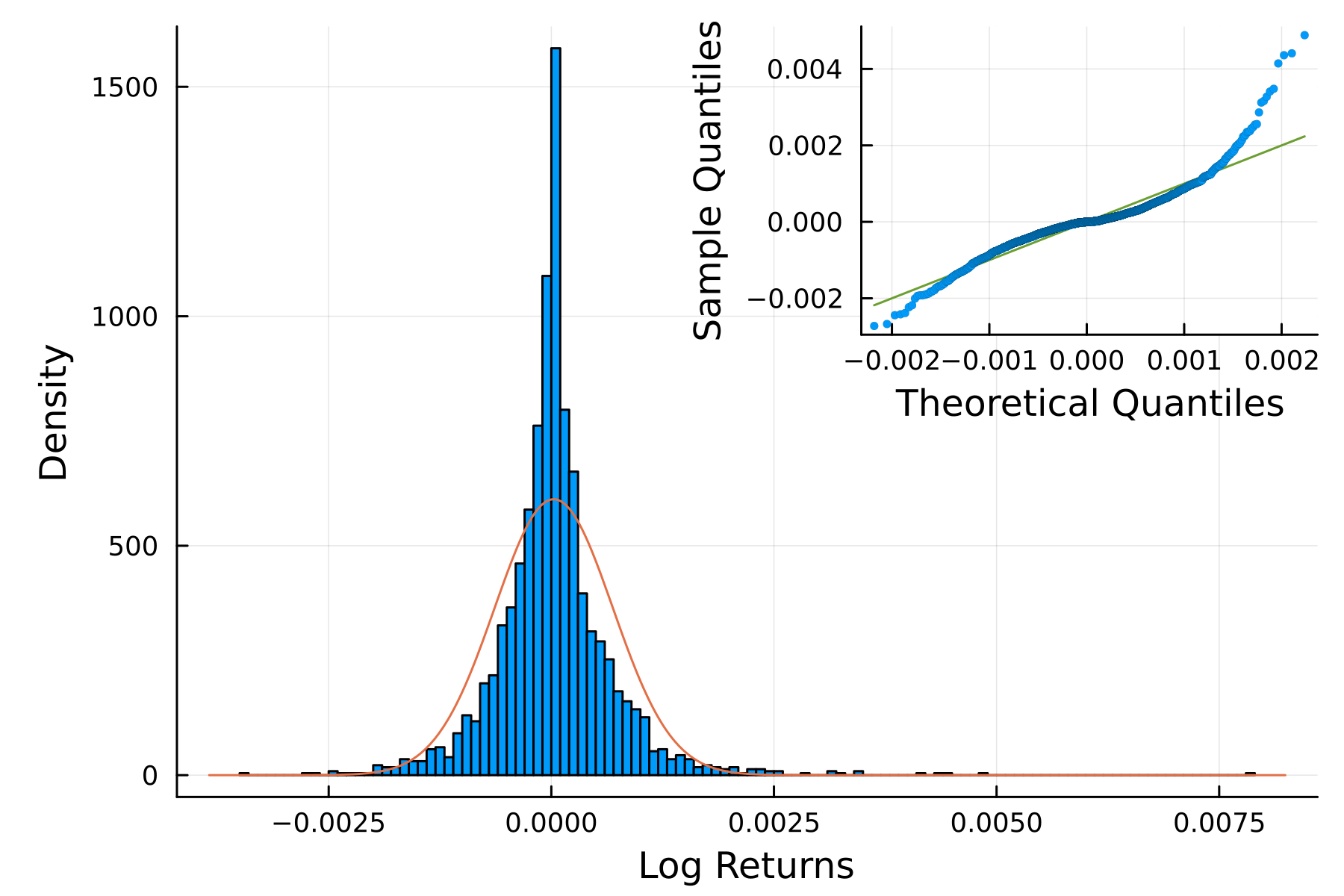}
    \caption{Empirical Returns Distribution \& QQ plot} \label{subfig:sf-emp-hist}
\end{subfigure}\hfill
\begin{subfigure}[t]{0.32\textwidth}
    \includegraphics[width=\textwidth]{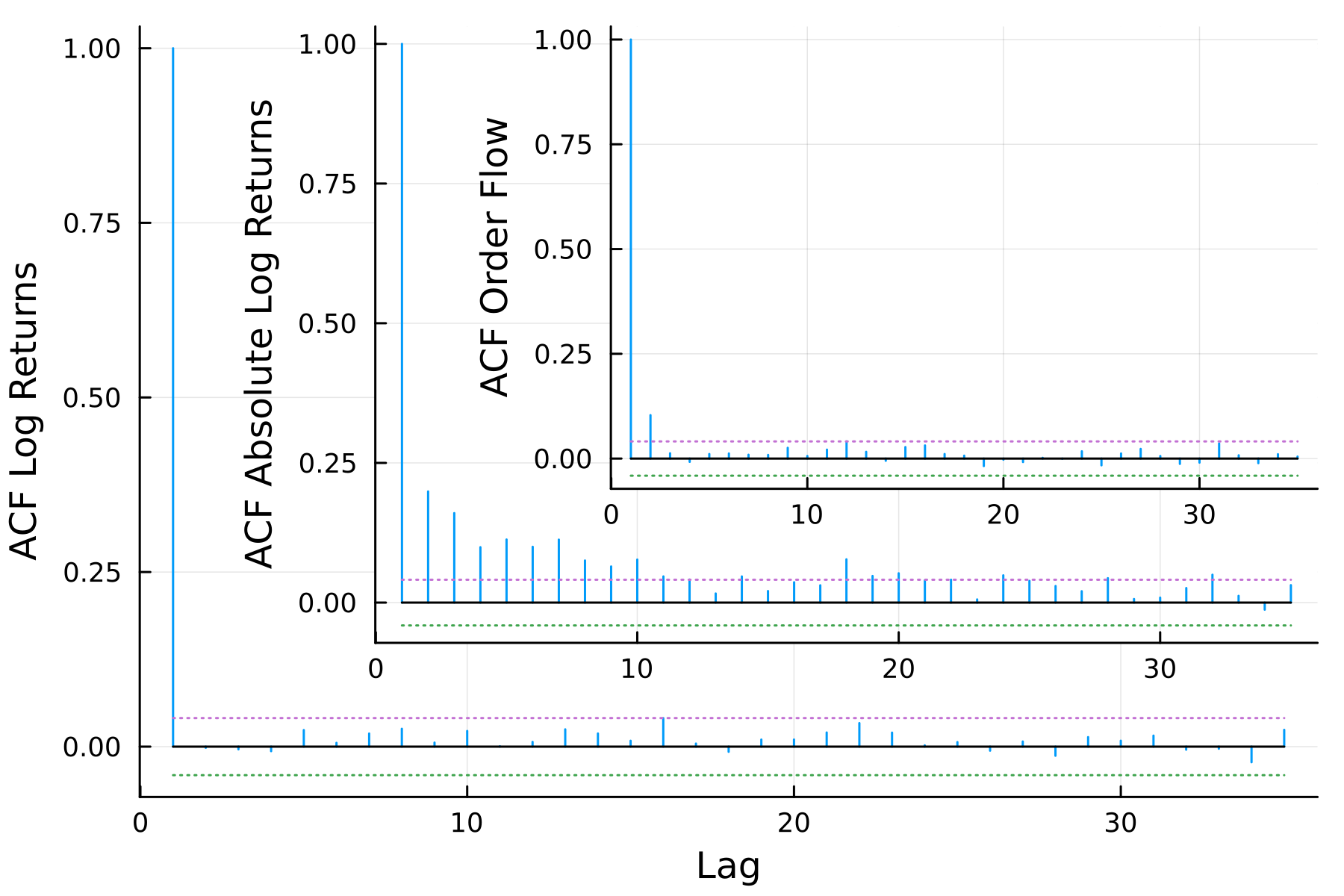}
    \caption{Empirical Autocorrelations} \label{subfig:sf-emp-acf}
\end{subfigure}
\par\bigskip
\begin{subfigure}[t]{0.32\textwidth}
    \includegraphics[width=\textwidth]{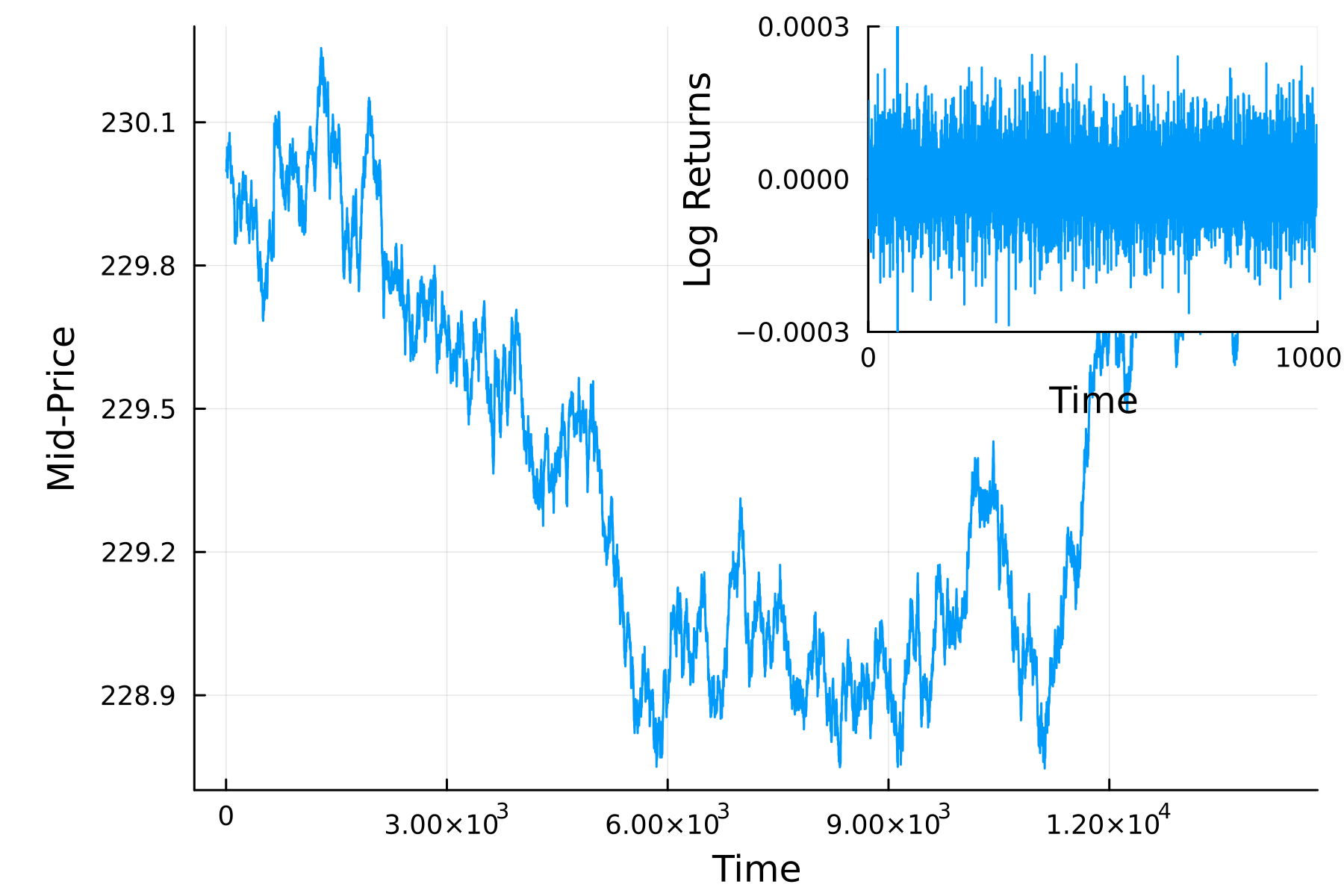}
    \caption{Non-uniform $\Delta t$ Price Path \& Returns} \label{subfig:sf-cal-non-unfiorm-price}
\end{subfigure}\hfill
\begin{subfigure}[t]{0.32\textwidth}
    \includegraphics[width=\textwidth]{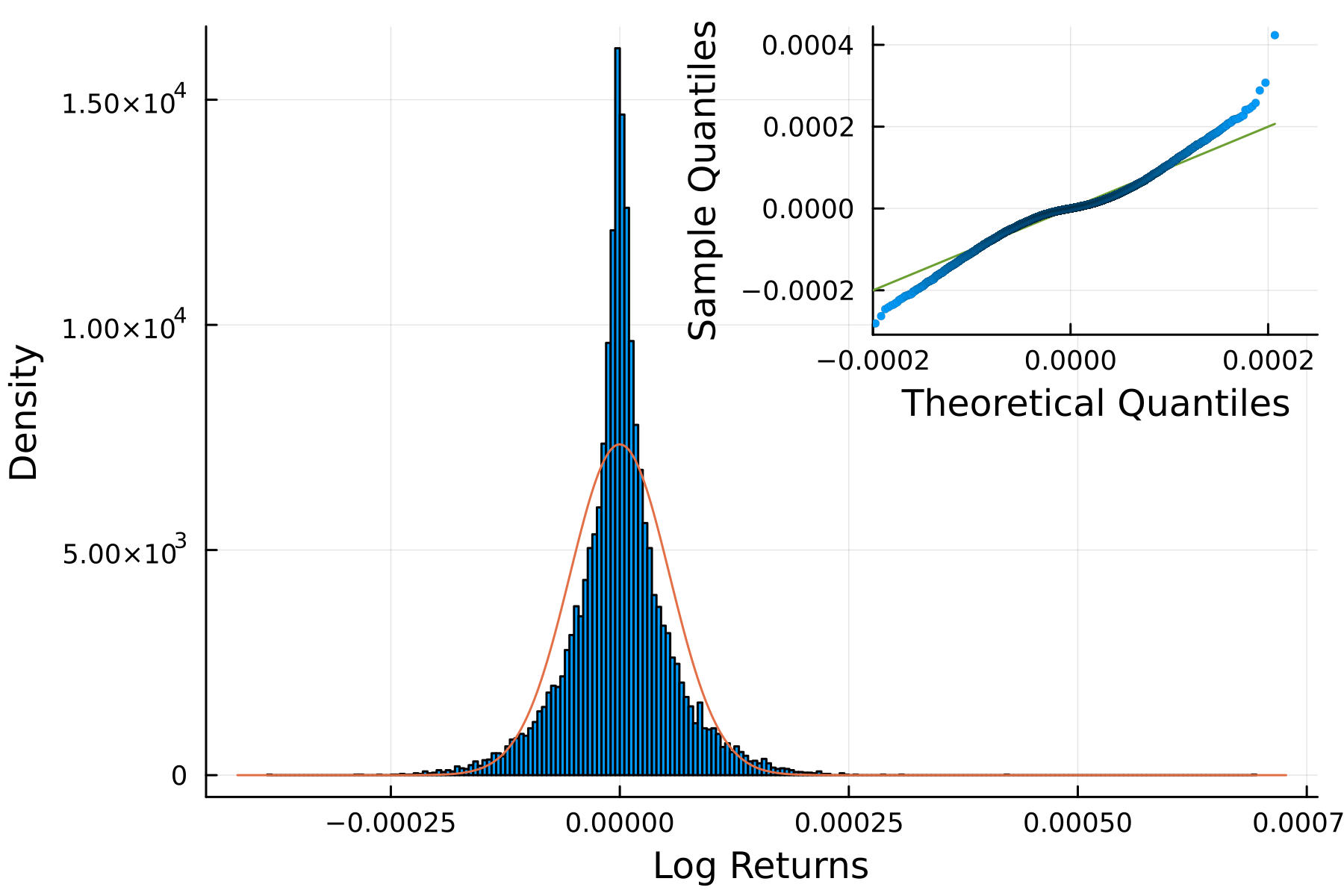}
    \caption{Non-uniform $\Delta t$ Returns Distribution \& QQ plot} \label{subfig:sf-cal-non-unfiorm-hist}
\end{subfigure}\hfill
\begin{subfigure}[t]{0.32\textwidth}
    \includegraphics[width=\textwidth]{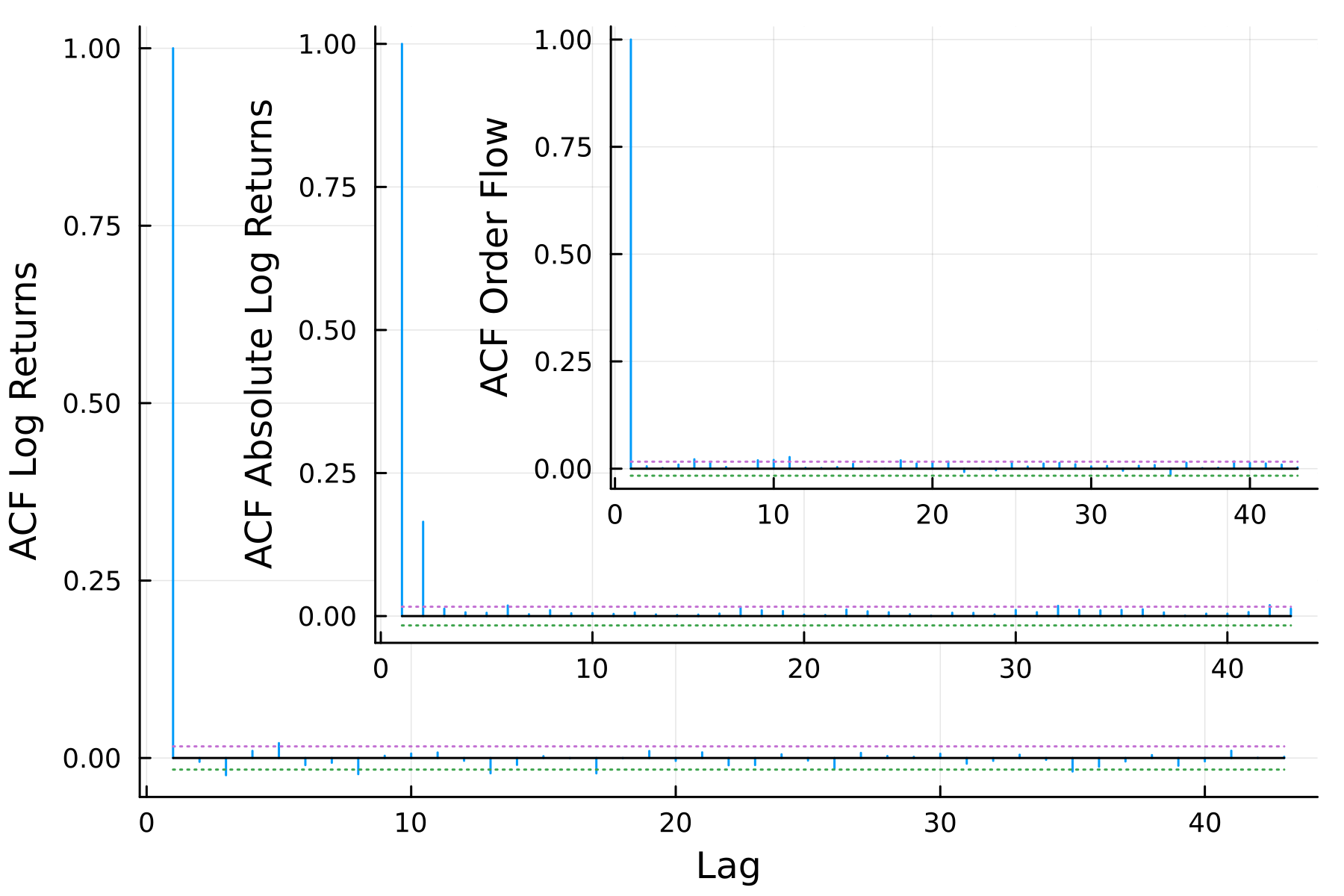}
    \caption{Non-uniform $\Delta t$ Autocorrelations} \label{subfig:sf-cal-non-unfiorm-acf}
\end{subfigure}
\caption{The stylised facts from the empirical data (figures \ref{subfig:sf-emp-price}, \ref{subfig:sf-emp-hist} \& \ref{subfig:sf-emp-acf}), and the data generated when using non-uniform $\Delta t$ (figures \ref{subfig:sf-cal-non-unfiorm-price}, \ref{subfig:sf-cal-non-unfiorm-hist} \& \ref{subfig:sf-cal-non-unfiorm-acf}) using the calibrated model parameters from Table \ref{tab:calibrated-parameters}. Our calibrated model produces a similar output of stylised facts as those shown in figures \ref{subfig:sf-emp-price}, \ref{subfig:sf-emp-hist} \& \ref{subfig:sf-emp-acf}. In terms of the price returns, Figure \ref{subfig:sf-cal-non-unfiorm-price} shows the returns and price curves. Figure \ref{subfig:sf-cal-non-unfiorm-hist} shows returns distributions and their expected departures from normality. We expect a rapid decay in the autocorrelations functions (ACF) for the returns, but not for that absolute value of the returns and order flow; Figure \ref{subfig:sf-cal-non-unfiorm-acf} shows a slow decay in the absolute value and order flow ACF but not as sustained as seen in the empirical case in Figure \ref{subfig:sf-emp-acf}.}
\label{fig:stylized-facts}
\end{figure*}

\section{Model Calibration} \label{sec:calibration} 

To improve the model parameters in Table \ref{tab:parameters} we calibrate them using empirical data. The parameters which dominate the model behaviour are found from a sensitivity analysis; these are found to be the diffusion rate of orders, the order cancellation rate, and the fractional derivative exponent. These three parameters are therefore considered to be free, and the same across all the coupled order books. This ensures that the only features driving the calibrated simulation parameters are shared, while the those related to the trading agents using the cross-coupling, and are unique to each order book from Equation [\ref{eq:ell-crosscoupling}], are fixed. The calibrated parameters and fixed parameters are given in Table \ref{tab:calibrated-parameters}. 

The key features these three free parameters have are the magnitude in variation of the confidence intervals for the diffusion rate $D_{\alpha}$, and the cancellation rate $\nu$; these show that weexpect high variation in these parameters across any simulation epoch. In contrast, the confidence intervals for derivative fraction $\alpha$ has less variation, and as such we have more confidence in its value and is expected to be more stable across epochs of real-world data. Table \ref{tab:calibrated-parameters} shows the calibrated variables used in the model with indicative confidence intervals using the method of \citet{jericevich2021simulation}.

\begin{table*}[t]
\centering
\begin{tabular}{cllllll}
 & & & &\multicolumn{3}{c}{Calibrated Values}\\ 
 \hline
Parameter & Description & Type & Range & $\Theta_{0.025}$ & $\Theta$ & $\Theta_{0.975}$ \\ 
 \hline
 $L$ & System length & Fixed & & & 200 & \\ 
 $M$ & Number of divisions & Fixed & & & 400  & \\ 
 $r$ & Probability of self jump & Fixed & & & 0.5 & \\ 
 $D_{\alpha}$ & Diffusion constant & Free & $[0, \infty)$ & -4.018 & 0.27 & 4.558 \\ 
 $\nu$ & Cancellations rate & Free & $[0, \infty)$ & -21.635 & 12.55 & 46.735 \\ 
 $\alpha$ & Fractional time & Free & $[0.4, 1]$ & 0.484 & 0.57 & 0.656 \\
 $p_{1,2}(0)$ & Initial prices & Fixed & & & 230 & \\ 
 $\lambda_{1,2}$ & Source terms intensities & Fixed & & & 1  & \\ 
 $\mu_{1,2}$ & Source terms rates & Fixed & & & 0.1 & \\ 
   $\Delta x$ & 
Change in $x$ ($\frac{L}{M}$) & Fixed & & & $0.5$ &\\ 
    $\Delta t$ & Change in $t$ (see Equation [\ref{eq:d}]) & Free & & & $0.2315$ & \\ 
 \hline
\end{tabular}
    \caption{The calibrated model parameters with description and bounds. When running the NMTA algorithm we found the calibrated values for $D_{\alpha}$, $\nu$ and $\alpha$ along with their confidence intervals. Both $D$ and $\nu$ have large confidence intervals which indicate a very high variability and as such less trust in the actual value, while $\alpha$ has less variability indicating more trust in the actual value. The value for $\Delta t$ has changed compared to Table \ref{tab:parameters} as a result of the change in $D_{\alpha}$. We use these parameters for the Epps effect in Figure \ref{fig:epps-calibrated} and the stylised facts in Figure \ref{fig:stylized-facts}.}  \label{tab:calibrated-parameters}
\end{table*}

\subsection{Calibration Data} \label{ssec:calibration-data}

When using the empirical data we needed to filter and prepare the data for analysis. This involved discarding data related to the opening and closing auctions, focusing only on trading activity that occurred between 9:00 and 16:50. Additionally, we observed that the first minute of trading produced unreliable data, so we decided to exclude it from our analysis to facilitate the model calibration.

To ensure the data was clean and relevant, we removed events associated with intraday volatility auctions and the impact of various futures close-outs. We also eliminated unwanted trades, such as after-hour trades (LT), corrections of the previous day's trades (LC), and auction uncrossing price trades (IP), focusing solely on automated trades (AT).

In financial markets, it's common for larger trades to be executed as a combination of smaller trades. This can lead to a single trade at a specific timestamp being split into multiple smaller trades, impacting the best bid and best ask prices as they fluctuate in the order book. The data may represent this event as multiple trades when it's actually a single trade. Additionally, some timestamps may have multiple quotes associated with them.

To address these complexities in the data, we conducted trade and quote compacting. This process involved modifying the data to better reflect the occurrence of a single trade and ensuring that there was only one quote per timestamp. When multiple quotes shared the same timestamp, we retained the most recent quote and removed the others. For trades with the same timestamp and the same order type, we calculated the aggregated trade volume for that timestamp and determined the price as the volume-weighted average price. This step was essential for cleaning and simplifying the data for our analysis.

Using the parameters from Table \ref{tab:parameters} in our model we are able to produce a price path for each of the two coupled order books for the non-uniform $\Delta t$ case and uniform $\Delta t$ case. We then use these price paths to produce Figure \ref{fig:epps-calibrated} which plots the correlations of the price paths $\rho$ against changes in time scale $\Delta t$ for non-uniform sampling. We further include the power density curve as an inset in the bottom right of both figures.

\subsection{Calibration Method} \label{ssec:calibration-method}

Given the model is simulating intraday trading, we follow the insights found in the agent-based modelling (ABM) literature. In the early stages of ABM research, parameters for the model were chosen manually to highlight that they could generate very specific statistical characteristics in the simulated data {\it i.e.} the empirically known stylised facts. 

However, there was no assessment to determine if the distribution of the generated data matched that of real empirical data, and no demonstrations of the extent to which parameter choices were pathologically chosen to unreasonably tune the model; where they would otherwise be unstable under small parameter changes (have large parameter sensitivity and degeneracies), or merely reflecting instability in the choice of objective functions in the presence of multiple similar local optima (objective function degeneracy). 

This implied that many of the early attempts in the ABM literature to claim generalisability of this or that insight were not statistically meaningful, and often did not always faithfully reflect the models themselves. This created a calibration challenge, where the model needed to be fine-tuned to better match real-world data \citep{fabretti2012, platt2017problem} in a more statistically believable manner \citep{platt2020}.

To address this calibration problem, researchers typically employ one of three common methods for financial ABMs: i.) Maximum Likelihood Estimation, ii.) Bayesian Inference, and iii.) Method-of-Moment using Simulated Minimum Distance (MM-SMD) \citep{platt2017problem}. We use the MM-SMD method because it is computationally efficient and does not require closed-form solutions. MM-SMD involves matching moments of the simulated data to those of empirical data, making it a practical choice for calibration. MM-SMD has known and well-understood limitations \citep{platt2020}. 

Despite these shortcomings (as an estimation method), MM-SMD remains a quick, straightforward, and reliable method for calibrating intraday trading models (as a calibration method). Particularly when there are more moments than parameters. This is why it is our chosen method of calibration. However, it is crucial to consider the calibrated parameters as indicative rather than robust (that is useful for inference) due to these inherent issues, and we use the same moments as used in the prior work that developed the simulation infrastructure \citep{jericevich2021simulation}. 

The estimation of these moments is carried out on the micro-price log returns, as opposed to mid-prices \cite{jericevich2021simulation}. As in prior work, the number $I$ of Monte Carlo replications is key to reducing the variance of the stochastic approximation of the objective function. We set $I = 5$ as it is more computationally practical, and noted that a number greater than $5$ is ideal to reduce the effects of randomness but compromises computational tractability with little apparent advantage in terms of the overall effectiveness of the calibration.  

As in prior work, we combine Threshold Accepting (TA) and Nelder-Mead (NM) algorithms (NMTA). In the NMTA algorithm, at each iteration, a decision is made to execute either a single step from the TA algorithm or a single step from the NM algorithm. The choice between a TA or NM algorithm step is probabilistic. During initial testing using the NMTA algorithm, we found that the values required bounds to observe better results in terms of the stylised facts and the Epps effect. The bounds used are as follows: $\nu \in [0, \infty)$, $D_{\alpha} \in [0, \infty)$, $\alpha \in [0.4, 1]$. We ran 100 iterations of the algorithm on each attempt at calibration. The values were found to be highly sensitive and convergence changed significantly during each attempted run of the NMTA algorithm. 

As such the final values can not conclusively be considered as the most optimal values in isolation from the noise sources -- however, this is well suited for the calibration use case explored here due to the computational efficiency of the method and its ability to extract of parameter set combinations in conjunction with random seeds that give good simulated path behaviour on the basis of sufficiently recovering some representation of the necessary stylised facts.

\section{The Epps effect} \label{sec:Epps}

\begin{figure}[th]
\centering
\includegraphics[width=0.49\textwidth]{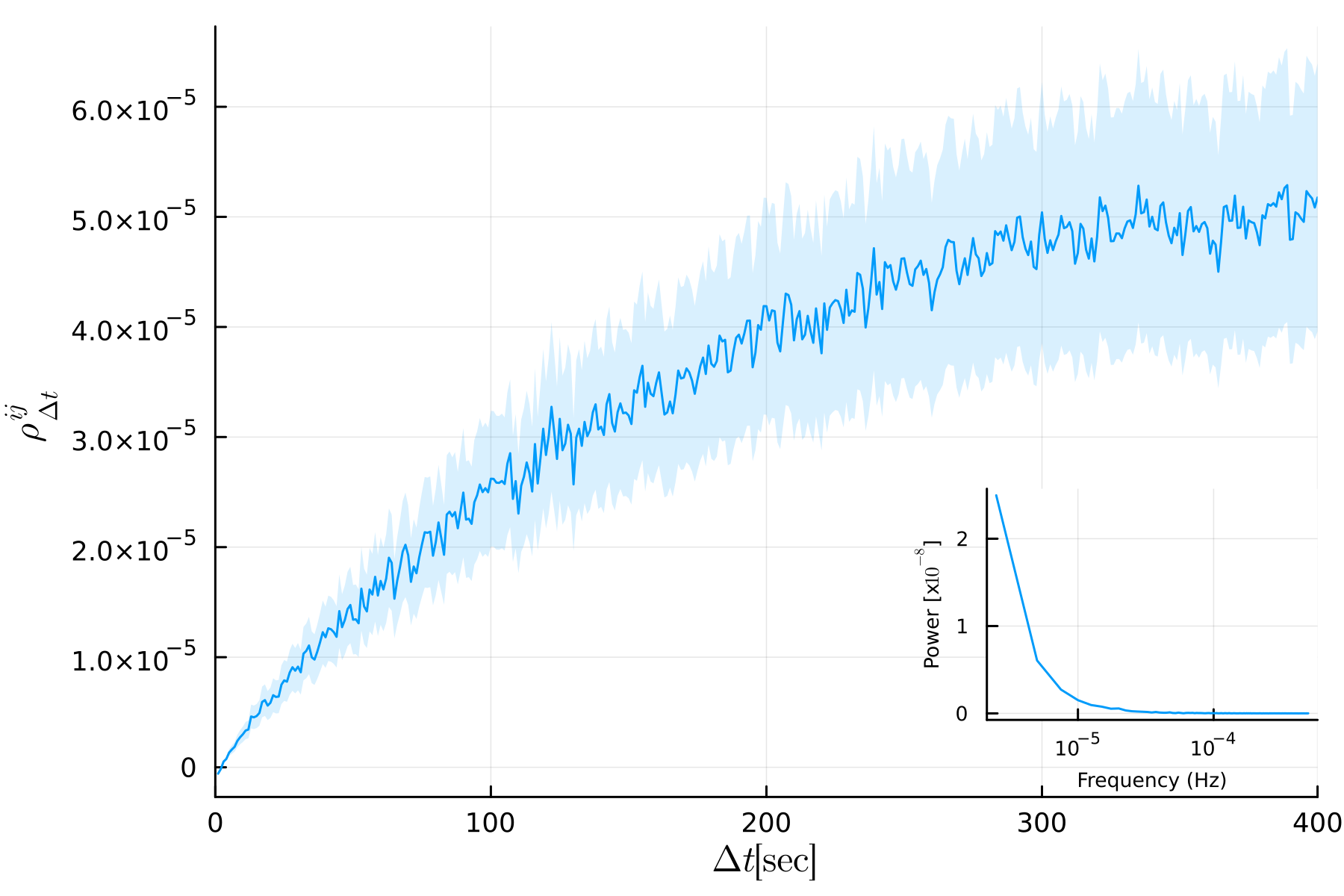}
\caption{Using the calibrated parameter values found in Table \ref{tab:calibrated-parameters} we generate the Epps effect for non-uniform $\Delta t$ with an accompanying power spectrum inset. The correlations of each figure have been averaged over 10 iterations. The associated price paths are in Figure \ref{fig:nonuniform-pricepaths} and these show the price paths generated using the coupled order book equation in Equation [\ref{eq:coupledRDequation}] with parameters defined in Table \ref{tab:parameters} when we couple two order books for the case of non-uniform $\Delta t$. }
\label{fig:epps-calibrated}
\end{figure}

\citet{Epps} showed that for very short time frames (such as minutes) correlated stocks show very little to no correlation. However, if you observe the correlated stocks over a longer time frame (such as hours or days) they tended toward their expected correlations. In other words, when you observe stock returns very frequently, they may seem less correlated, but when you aggregate the data over longer time intervals, the correlations become more pronounced. This phenomenon has been corroborated in other studies on equities \citep{bonanno2000, zebedee2009, tumminello2006, mastromatteo2010} and foreign exchange markets \citep{muthuswamy}.

There are three sources which appear to contribute to the Epps effect - asynchrony, lead-lag and tick-size \citep{CHANG2021126329}, However, these all seem to be artefacts of the fundamentally discrete event nature of real financial markets \citep{chang2021using}. These sources were explored by others. \citet{munnix2010} explored the effect that the tick-size has on the Epps effect. \citet{reno2001} investigated asynchrony when there is a lead-lag present. \citet{precup2007} explored varying the levels of asynchrony and showed that this resulted in differently behaving Epps effects. An analytical representation for the Epps effect was shown in \citet{toth2007, toth2009} and extended by \citet{mastromatteo2010} to differentiate between the effects of lead-lag and asynchrony.

\citet{chang2021using} speculate that the Epps effect should best be understood as an emergent property under synchronised averaging of observations sampled from  discrete financial markets market events. They questioned whether infinitesimal diffusion processes with a unique global time are an appropriate underlying latent model representation.  

Here the system comprises discrete events that do not naturally align in a unique global trading time, when a trading agent couples the two asynchronous and discrete order books trying to exploit price differences, then there are events that cannot be uniquely aligned when compared cross-sectionally. This means that correlations can only emerge on averaging scales sufficiently long that there are enough events in the discrete sample periods to meaningful allow estimation. This implies that there are no meaningful latent correlations -- correlations are an extrinsic property. Correlations are the result of an averaging procedure that is separate from the data generating process and merely represent the average impact of trading at low frequencies without being an intrinsic property of the order book itself.  

To show this, we will first need to estimate the correlations. We implement the non-uniform fast Fourier transform (NUFFT) using the Dirichlet basis kernel with Fast Gaussian Gridding (FGG) as described in \citet{chang2020}. The reasons for using this method are: NUFFT methods work well with data that is asynchronous, discrete and event-driven as is the case with our data, the Dirichlet kernel has been shown to plausibly recover the theoretical Epps
curve and FGG is computationally efficient. Briefly, the algorithm first convolves nonuniform source points on the over-sampled grid, carries out a Discrete Fourier Transform (DFT) on the over-sampled grid with the standard FFT, and then deconvolves the convolution in the Fourier space.

Using the values from Table \ref{tab:calibrated-parameters} we generate the figures in Figure \ref{fig:epps-calibrated}. Both figures show the correlations over different time scales with the power spectrum as an inset in the bottom right-hand corner of the figure. For Figure \ref{fig:epps-calibrated} we use non-uniform sampling. We see the emergence of the Epps effect. Further, we see the power spectrum graph decays.

The stylised facts for the empirical data along with both the non-uniform and uniform $\Delta t$ cases are shown in Figure \ref{fig:stylized-facts}. Each of the cases has three accompanying stylised facts figures. Each of the three cases (i.e. empirical, non-uniform and uniform $\Delta t$) going from the left-most figure to right most has a price curve with a log-returns inset in the top right-hand corner, followed by the log returns distribution with a QQ-plot of the returns in the top right-hand corner and a figure with three autocorrelation curves (ACF) for the log returns, absolute log returns and order flow. Figures \ref{subfig:sf-emp-price}, \ref{subfig:sf-emp-hist} and \ref{subfig:sf-emp-acf} show the stylised facts for the empirical results. The output of these figures is what we would like our model to reproduce. 

Figures \ref{subfig:sf-cal-non-unfiorm-price}, \ref{subfig:sf-cal-non-unfiorm-hist} and \ref{subfig:sf-cal-non-unfiorm-acf} are the stylised facts for the non-uniform $\Delta t$ case. We see a realistic price curve with noisy log returns in Figure \ref{subfig:sf-cal-non-unfiorm-price}. We see deviations from a normal distribution of log returns, with additional tail-events and concentration around the mean; an almost S-like QQ plot in Figure \ref{subfig:sf-cal-non-unfiorm-hist}. We only see an initial autocorrelation decay for the ACF of absolute log returns and no decay for the order flow in Figure \ref{subfig:sf-cal-non-unfiorm-acf}.

\subsection{Results Discussion} \label{ssec:null-case}

Using the parameters from Table \ref{tab:parameters} in our model to produce price paths for the two coupled order books for a variety of different non-uniform $\Delta t$ (and uniform sampling) cases. We observed the correlations over varying changes in time scale ($\Delta t$) between two coupled Brownian motions. We use this to confirm that if our model does produce an Epps effect it is not due to how we compute our correlations. Using empirical data from \citet{Chang2020-data} to observe the correlations between two banking stocks listed on the JSE, namely, Standard Bank (SBK) and Nedbank (NED). We confirmed that we are able to reproduce this finding using our method of computing correlations as the correlations form an Epps effect over changes in time scale ($\Delta t$).

We then explored the effect of changing $\Delta{t}$ on the output of the Epps of effect. We considered the effect of increasing $\Delta{t}$ while keeping a constant $\Delta{x} = \frac{1}{2}$. We accomplish this by using Equation [\ref{eq:d}] and changing the value of $D_{\alpha}$, the diffusion rate. We choose to hold $r$, the probability of a self-jump, constant. We started with $\Delta{t} = 0.0625$ ($D_{\alpha} = 1.0$), then next has $\Delta{t} = 0.0781$ ($D_{\alpha} = 0.8$) followed by $\Delta{t} = 0.125$ ($D_{\alpha} = 0.5$) and finally we end with $\Delta{t} = 0.2083$ ($D_{\alpha} = 0.3$). The Epps effect was evident but weakened with increasing sampling.   

The effect of changing $\Delta{x}$ while keeping a constant $\Delta{t} = 0.03$ on the emergence of the Epps effect was then considered. We started with $\Delta{x} = 0.25$ then $\Delta{x} = 0.33$ followed by $\Delta{x} = 0.5$ and finally ending with $\Delta{x} = 0.75$. The Epps effect is again, unsurprisingly, shown to have a weakened overall equilibrium correlation as the sampling scale increases.

The effect of changing $\nu$ on the Epps effect when we use non-uniform $\Delta t$ combined with the parameters from Table \ref{tab:parameters}. The Epps effect for values of $\nu$ starting at $\nu = 1$ and incrementing by $3$ until we reach a value of $\nu = 16$. We found that at $\nu = 1$ we are able to recover the Epps effect without any numerical artefacts (drops in correlation at specific frequencies multiples of the sampling rate). However, as $\nu$ increases we started to see spikes in correlation that we have previously seen, particularly in the uniformly sampled case, that are known numerical sampling artefacts. There appears to be a positive correlation between the magnitude of $\nu$ and the magnitude of the drop in correlation.

We checked the methods use here by first measuring the observed correlations over varying changes in time scale $\Delta t$ between two coupled Brownian motions. We use this to confirm that if our model does produce an Epps effect it is not due to how we compute our correlations. We report that we were able to recover the required constant and zero correlations as a function of the increasing time changes confirming that there is no meaningful measurement method induced correlations in our simulation configuration. 

We the use empirical data from \citet{Chang2020-data} to observe the correlations between two banking stocks listed on the JSE, namely, Standard Bank (SBK) and Nedbank (NED). These are an example of correlated shares which produce an Epps effect \citep{CHANG2021126329}. We confirm that we are able to reproduce this finding using our method of computing correlations as the correlations form an Epps effect over changes in time scale $\Delta t$ as observed in the prior work. At large enough sampling time scales we find no Epps effect and a decreasing trend in correlation $\rho$ as a function of the time lag.

The effect of changing the cancellation rate $\nu$ on the Epps effect, when we use non-uniform $\Delta t$ combined with the parameters from Table \ref{tab:parameters} was then considered. We considered the Epps effect for values of $\nu$ starting at $\nu = 1$ and incrementing by $3$ until we reach a value of $\nu = 16$. At $\nu = 1$ we are able to recover the Epps effect without any numerical artefacts in correlation. As $\nu$ increases we observed drops in the correlation that are due to aliasing in the numerical scheme; this is a problem known in the uniformly sampled numerical scheme. There appears to be a positive correlation between the magnitude of $\nu$ and the magnitude of the drop in correlation due to numerical artefacts. 

As $\nu$ is increased the drops in correlation increase in size due to numerical noise. This implies that the fits are strongly dependent on the parameters not only because of parameters being tuned to the physical properties of the data but also because under calibration there are combinations of parameters that resolve data dynamics as balanced against numerical noise and sample noise from the Monte Carlo steps. This confirms that these methods are not yet fit for estimation work, but can be calibrated to recover results for simulation work. This needs significantly more diagnostic work to better understand the complex interplay between the numerical scheme, its stability, and the models ability to recover various stylised facts under calibration. 

\section{Conclusions}\label{sec:6}

Here we demonstrate how a relatively simple microscopic coupled order book model can generate the Epps effect \cite{Epps}. This is an empirical characteristic found in correlated financial assets, where the correlations start to vanish when they are measured over decreasing time scales. The model for each of the coupled order books is given as a reaction-diffusion system as in Equation [\ref{eq:coupledRDequation}]. This has been shown to be equivalent in the diffusion limit to a discrete model that can be numerically simulated 
using a convenient stochastic update equation \cite{Angstmann2016,ANGSTMANN2016508,AHJM2016,Diana2023} (Equation [\ref{eq:PDECoupledUpdateEquation}]).

Figure \ref{fig:nonuniform-pricepaths} is an example of price paths generated when using Equation [\ref{eq:PDECoupledUpdateEquation}] with non-uniform sampling applied to two-coupled order books. Figure \ref{fig:price-dynamics-non-uniform} then showed time snapshots of what occurs when there is an order book shock in. 

To demonstrate the Epps effect we implement a non-uniform fast Fourier transform using the Dirichlet basis kernel with fast
Gaussian gridding \citep{chang2020} to estimate correlations between the price paths for different time scales. We first considered a null case to confirm that our estimation scheme does not obscure our results and then show that the Epps effect emerges for empirical data.

Thereafter we confirmed that the coupled price paths do produce the Epps effect for both the non-uniform time sample increments $\Delta t$, and uniform $\Delta t$ sampling cases. In the case of uniform sampling, there are periodic drops in correlation. To better understand the cause of these drops in correlation we completed several simulation experiments and report these. We found that increasing the sampling time $\Delta t$ leads to an increase in the magnitude of the drops in correlation. We found a similar result for an increase in log-price grid size $\Delta x$, however, we saturate the model and lose the Epps effect when the log-price grid size is too large. 

We varied the cancellation rate $\nu$ for the non-uniform and uniform time sampling to show that increasing the cancellation rate resulted in increases in the size of the periodic drops in correlation, with the uniform sampling case showing larger decreases compared with the non-uniform sampling case, and with significant numeric instability. We found that a cancellation rate value of $\nu$ close to $1$, given the other parameters are held constant, results in no drops in correlation for both cases.

Finally, we use empirical data in Section \ref{sec:calibration}  and the method of moments combined with a Nelder-Mead method with threshold acceptance optimisation algorithm to calibrate the model parameters, these are shown in Table \ref{tab:calibrated-parameters}. We found that the calibrated values for $D_{\alpha} = 0.27$ and $\nu = 12.55$ and a large confidence interval indicating that their values varied across replications, whereas the value of $\alpha = 0.57$ was found with some confidence. We found that this combination of parameters produced a desired Epps effect for the non-uniform time sampling (Figure \ref{fig:epps-calibrated}), which was surprising given the large order cancellation rate.

We further compared the stylised facts produced by both cases to the empirical stylised facts we wish to reproduce in Figure \ref{fig:stylized-facts}. The non-uniform case performed well and resulted in stylised facts which closely resemble the empirical effects, more so than those associated with the uniform sampling. However, improvements can still be made with regard to the autocorrelations. In conclusion, we were able to reproduce the emergence of correlations in the system of coupled order books using a mechanistic pair-trader in combination with other stylised facts that corresponded reasonably well to those found in real-world market data.

\section*{Acknowledgements}

We thank Chris Angstmann and Byron Jacobs for many helpful and interesting conversations.

\section*{Data and Software}
The code base used here extends that developed by Derick Diana \cite{Diana2023}. The coupled-book development is described in \citet{bauer2024-thesis}; the Julia code used for the simulations is archived on ZivaHub \citep{bauer2024-zivahub} and is available in the \texttt{InteractingLOBs.jl} GitHub repository \citep{bauer2024-github}.

\bibliography{CoupledOB}

\end{document}